\documentclass[runningheads]{llncs}

\usepackage{graphicx} 
\usepackage{braket}
\usepackage{tikz}
\usepackage{amsmath}
\usetikzlibrary{quantikz2}
\DeclareMathOperator{\argmin}{argmin}

\begin{document}

\title{Quantum Computing in Logistics and Supply Chain Management - an Overview}
\titlerunning{Quantum Computing in Logistics and Supply Chain Management} 

\author{Frank Phillipson}
%
\authorrunning{F. Phillipson}

\institute{Maastricht University, Maastricht, The Netherlands \\
TNO, The Hague, The Netherlands \\
\email{f.phillipson@maastrichtuniversity.nl}}


\maketitle

\begin{abstract}
The work explores the integration of quantum computing into logistics and supply chain management, emphasising its potential for use in complex optimisation problems. The discussion introduces quantum computing principles, focusing on quantum annealing and gate-based quantum computing, with the Quantum Approximate Optimisation Algorithm and Quantum Annealing as key algorithmic approaches.

The paper provides an overview of quantum approaches to routing, logistic network design, fleet maintenance, cargo loading, prediction, and scheduling problems. Notably, most solutions in the literature are hybrid, combining quantum and classical computing. The conclusion highlights the early stage of quantum computing, emphasising its potential impact on logistics and supply chain optimisation. In the final overview, the literature is categorised, identifying the dominance of quantum annealing and highlighting the need for more research in prediction and machine learning. The consensus is that quantum computing has great potential but faces current hardware limitations, necessitating further advancements for practical implementation.

\end{abstract}

\section{Introduction}
\begin{figure}
    \centering
    \includegraphics[width=8cm]{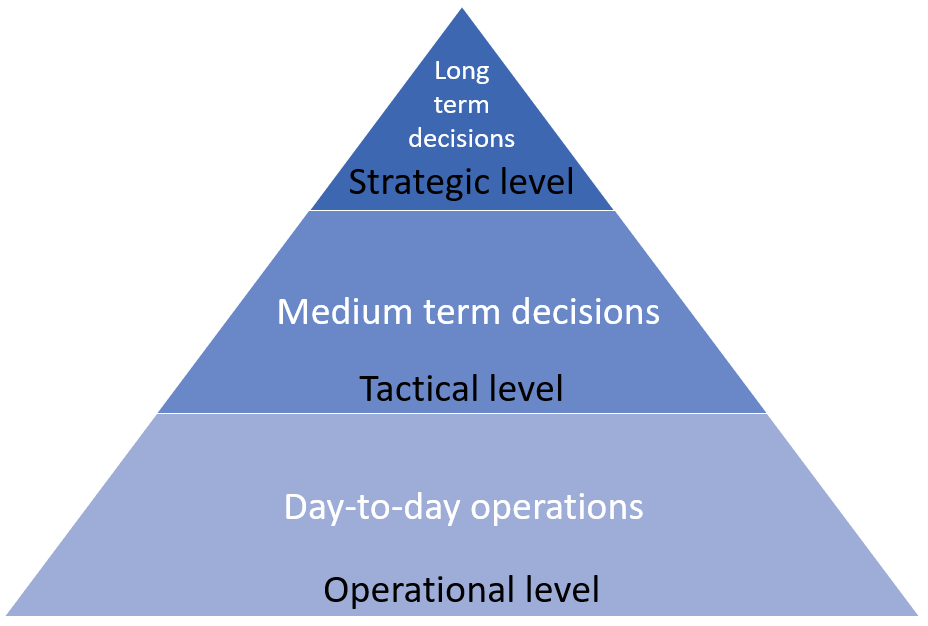}
    \caption{Three levels of problems.}
    \label{pic:pyramide}
\end{figure}

Quantitative optimisation is integral to logistics and supply chain management, driving efficiency, cost reduction, and overall performance improvement. Various challenges underscore the need for this field. These kinds of challenges and the corresponding optimisation problems are often divided into multiple levels, such as operational, tactical and strategic \cite{brouer2018optimisation,de2017framework}, see Fig. ~\ref{pic:pyramide}. On an operational level, there is efficient route planning, as seen in vehicle routing problems (VRP), minimising transportation costs while meeting demand and adhering to constraints. Problems like stowage and cargo loading also need to be solved on an operational level. Next, in inventory management, a balance is sought between over-stocking and under-stocking, employing models like Economic Order Quantity (EOQ) for optimal order quantities. Production schedule optimisation addresses job scheduling, employee and machine scheduling, and production planning, minimising lead times and costs. Lastly, accurate demand forecasting is needed to optimise inventory, production, and transportation planning. On a tactical level, we have problems like fleet deployment, timetabling, and hub site design. On a strategic level, decisions on facility location, fleet size and mix, and network design, impact logistics costs. 

In addition to these pure optimisation problems, there are two issues we also have to address in quantitative optimisation in this field. First, supply chain objectives often conflict, requiring multi-objective optimisation for informed decisions. Supplier selection involves balancing factors like cost, quality, lead times, and reliability. Risk management minimises disruptions from natural disasters or political events, ensuring operational continuity.
Second, in logistics, we have to cope with the dynamic behaviour of the system. Dynamic optimisation adapts to real-time changes, crucial in a dynamic supply chain environment. Green supply chain optimisation focuses on sustainability, reducing environmental impact through carbon emission and energy usage minimisation.

These issues make decision-making processes even more complex, considering various variables and constraints, ensuring a smooth flow of goods while maximising operational efficiency and minimising costs. Advanced techniques, including mathematical modelling and metaheuristics, are commonly used to tackle these challenges, which are computationally hard in most cases.\\

Currently, there is a burgeoning computing paradigm known as quantum computing, which holds the potential to transform the contemporary computing landscape. Quantum computers, the devices central to this paradigm, leverage principles from quantum mechanics to achieve computational speeds and efficiencies that surpass the inherent capabilities of current classical computers for specific problems. The notable advancements attributable to quantum computing include a quadratic improvement in searching unstructured databases~\cite{Grover_1996} and an exponential enhancement in integer factorisation~\cite{Shor_1994,Shor_1997}. Beyond these examples, quantum computing is expected to provide advantages across various domains, such as simulating chemical processes~\cite{Cao2019}, solving (or approximating) optimisation problems~\cite{jaderberg2022solving,Symons_2023}, and advancing quantum machine learning~\cite{phillipson2020quantum,schuld2021machine}.

Quantum computing also has the potential to revolutionise the field of computational logistics and supply chain management by offering advantages in solving complex optimisation problems better of faster. Some of these problems include solving combinatorial optimisation problems, which are prevalent in logistics as we saw before: vehicle routing, facility location, and network design problems, which involve finding the best routes, locations, and structures to minimise costs and improve efficiency, and scheduling problems, such as production scheduling, job sequencing, and workforce management. Note however that their ability to really solve NP-hard problems efficiently remains uncertain. Many combinatorial optimization problems, such as the Travelling Salesman Problem or Integer Programming, are NP-hard, meaning that no known polynomial-time algorithm exists for them on classical computers. Quantum computing operates within the complexity class BQP (Bounded-Error Quantum Polynomial Time), which includes problems that quantum computers can solve efficiently with high probability. However, NP-hard problems are generally believed to be outside BQP \cite{czerwinski2023np}, suggesting that quantum computers may not provide exponential speedups for solving them exactly. However, quantum (assisted) heuristics may offer advantages in finding approximate solutions faster than classical methods.

The Quantum Technology and Application Consortium (QUTAC) \cite{bayerstadler2021industry} also acknowledges that optimisation and simulation problems are prevalent in the domain of production and logistics across industries, such as manufacturing, chemical and pharmaceutical production, insurance, and technology. They state that real-world problems often involve numerous variables and constraints that classical algorithms struggle to address effectively, and that quantum optimisation approaches, such as quantum annealing (QA) and hybrid algorithms like the Quantum Approximate Optimisation Algorithm (QAOA), are promising for their potential to provide higher-quality solutions and faster solution times. Yarkoni et al. \cite{yarkoni2022quantum} also recognise scheduling and logistics as an important application area of quantum computing in their overview of industry applications.

By reducing computation time significantly, quantum computing can help model and optimise the supply chain in real time, making it (more) resilient to disruptions, by finding alternative routes and sources during unexpected events, thus minimising the impact of disruptions. Quantum computers are also expected to efficiently process and analyse vast amounts of data in the future, which is essential for demand forecasting, inventory management, and making real-time decisions in the supply chain. Lastly, quantum algorithms can tackle multi-objective optimisation problems by simultaneously considering multiple conflicting objectives. This is beneficial for balancing cost reduction, service level improvement, and environmental sustainability, which is a great benefit for the two issues mentioned before.

While quantum computing offers great promise for addressing logistics and supply chain optimisation problems, it is important to note that the technology is still in its early stages, and practical, large-scale implementations are limited. However, quantum hardware is evolving rapidly, and researchers are working on developing both the software stack needed and quantum algorithms for specific problems. Quantum computing-based approaches specifically tailored to logistics and supply chain management are also being studied and developed. A combination of classical and quantum computing techniques may be the most practical approach for solving logistics and supply chain challenges based on the current state of quantum computing hardware. This is reflected in literature as this work shows. 

To get grip on this, this paper provides an overview of over 80 published papers on this topic, both in detail and in the form of a compact overview. Next, conclusions from this overview are drawn, and blind spots are recognised for further research.
We continue the paper in Section \ref{sec:quin} with an introduction to quantum computing and optimisation. Next, in Section \ref{sec:overview}, we will dive into the various areas of quantitative optimisation within the logistics and supply chain world and provide an overview of the quantum approaches for the problems in those areas. As indicated, we will end with a concise summary and recommendations for further research.

\section{Quantum computing and optimisation} \label{sec:quin}
We can currently distinguish two paradigms in quantum computing: \textbf{digital} or gate-based computers/computing (GBC) and \textbf{analogue} quantum computing, of which quantum annealers are an important example. GBC is most similar in operation to the current generation of computers. They are capable of performing operations (gate operations, such as AND, OR) on specific qubits or on multiple qubits simultaneously. This allows for actual programming, which is often visualised through circuit diagrams. The QAs, on the other hand, are single-purpose machines. QA started with the work of Kadowaki and Nishimori \cite{kadowaki1998quantum}. QA can basically do one thing only: find the minimum value of a specific function. This function is encoded in the qubits, after which a quantum mechanical evolution leads to a solution that minimises the energy. In this section, we present GBC and QA in more detail and explain how they can be used in quantitative optimisation. First, we introduce some fundamental concepts in quantum computing.

\subsection{Fundamental Concepts}
Superposition, entanglement, interference, and tunnelling are fundamental concepts in quantum mechanics that underpin the power of quantum computing. Superposition refers to the ability of a quantum system to exist in multiple states simultaneously. Quantum computing uses qubits, or quantum bits. Unlike classical bits, which can only represent either 0 or 1, a qubit can be in a superposition of both states simultaneously. Mathematically, a qubit can be represented as a linear combination of the basis states $\ket{0}$ and $\ket{1}$, often denoted as $\alpha \ket{0} + \beta \ket{1}$, where $\alpha$ and $\beta$ are complex numbers that describe the probability amplitudes of each state. This notation is further explained in Section \ref{sec:GBC}. When a measurement is made, the qubit collapses into one of the basis states with a probability determined by the square of the amplitudes, so it should hold that $|\alpha|^2+|\beta|^2=1$. Quantum parallelism is related to superposition. Quantum computational advantage is achieved by leveraging superposition to perform parallel computations on multiple qubits. 

Entanglement describes a strong correlation that can exist between two or more qubits. When qubits become entangled, the state of one qubit becomes intrinsically linked to the state of the other qubits. This entanglement persists even if the qubits are physically separated. Entangled qubits can exhibit highly non-classical behaviour and can be used to perform quantum operations that are not possible with classical systems. Superposition and entanglement are fundamental resources in quantum computing, enabling the execution of powerful quantum algorithms and the potential for exponential computational speedup.

Quantum interference is the third fundamental phenomenon in quantum mechanics, where the amplitudes of different quantum states combine in such a way that they can reinforce or cancel each other out. It occurs when two or more quantum states, such as wave functions or probability distributions, overlap and interact with each other. Quantum interference is best understood through the concept of superposition. When superposed states interfere, the resulting probability distribution is not simply the sum of the individual probabilities. Instead, it depends on the relative phase of the states.

Lastly, quantum tunnelling is a phenomenon in quantum mechanics where a particle can pass through a potential barrier even when its energy is lower than the energy of the barrier. In classical physics, if a particle does not have enough energy to overcome a barrier, it would be reflected back or stopped by the barrier. However, in the quantum realm, particles such as electrons and protons can exhibit wave-like behaviour and have a non-zero probability of ``tunnelling" through the barrier.

\subsection{Gate based quantum computing} \label{sec:GBC}
The first paradigm is digital or gate-based quantum computing. GBC relies more directly on qubits and gate operations. As said before, unlike classical bits, which can only represent 0 or 1, qubits can exist in a superposition of states, simultaneously representing multiple values. Through the application of quantum logic gates, which are analogous to classical logic gates, quantum computations are executed. These gates manipulate the quantum states of qubits, enabling operations such as superposition, entanglement, and interference. By leveraging a sequence of carefully crafted gate operations, quantum algorithms can be executed, allowing for the solution of problems that are intractable for classical computers. The precise control and manipulation of qubits, as well as the mitigation of errors due to decoherence, pose significant challenges in the practical implementation of gate-based quantum computing systems.
The qubit states can be represented by matrices (vectors):
\begin{equation}
\ket{\psi}=\ket{0}=\binom{1}{0} \textrm{ and } \ket{\psi}=\ket{1}=\binom{0}{1}.    
\end{equation}

This means that a one-qubit quantum gate can be depicted as a unitary operator acting on a single qubit, which is represented as a two-dimensional system. A quantum gate acting on $n$ qubits is represented by a $2^n \times 2^n$ matrix. The state vector $\ket{\psi}$ belongs to a Hilbert space $\mathcal{H}$, called the state space with vectors of length 1, using complex numbers.
This state is given in Dirac notation: a quantum object state is represented by $\ket{\psi}$, the \textit{ket} of quantum state $\psi$. The \textit{bra} of the same state vector, represented by $\bra{\psi}$ is the conjugate transpose of the ket. If the vector contains only real numbers, the conjugate transpose is the same as the regular transpose, for a vector with complex entries, the conjugate transpose replaces each complex entry with its complex conjugate\footnote{if you have a complex number $z = a + bi$, where $a$ is the real part and $b$ is the imaginary part, then the complex conjugate of $z$, denoted as $\bar{z}$, is defined as $\bar{z} = a - bi.$} and then transposes the resulting matrix. Quantum programming languages can be used to construct and apply quantum circuits to quantum hardware. Examples of such languages are PyQuil \cite{koch2019introduction}, QCL \cite{omer2005classical} and Q\# \cite{tolba2013q}. These quantum programming languages only focus on that specific part of the quantum software stack \cite{van2019vision,piattini2020talavera}, whilst tools for other layers are also in development.

\subsection{The QUBO}
Many approaches on quantum computer devices that are related to optimisation use the QUBO, Quadratic Unconstrained Binary Optimisation, formulation as standardised input. This QUBO, or the equivalent Ising formulation, represents the function that can be minimised by the quantum algorithm. The Ising formulation defines the energy of electron spins via a Hamiltonian and enables it to minimise this energy. The QUBO formulation is its binary representation and is often used for combinatorial optimisation:
\begin{equation}
\text{QUBO:} \min \mathbf{x}^T Q\mathbf{x}\end{equation} where $\mathbf{x}$
is a vector with binary decision variables and $Q$ is a square matrix with constant values. Many (constrained) combinatorial optimisation problems can easily be described in a QUBO formulation \cite{glover2019quantum}. Take for example the problem
\begin{equation}
\min  \mathbf{c}^T \mathbf{x} \textrm{, under the constraints } A\mathbf{x}=\mathbf{b},
\end{equation}
then we can place the conditions in the objective function using a penalty $\lambda$:
\begin{equation}
\min  \mathbf{c}^T \mathbf{x}+\lambda(A\mathbf{x}-\mathbf{b})^T (A\mathbf{x}-\mathbf{b}). \end{equation}
If we define $P=I\mathbf{c}$ being the matrix having the values of $\mathbf{c}$ on the diagonal, then this equals:
\begin{equation}
\argmin  \mathbf{x}^T P\mathbf{x}+\lambda(A\mathbf{x}-\mathbf{b})^T (A\mathbf{x}-\mathbf{b})=\mathbf{x}^T P\mathbf{x}+\mathbf{x}^T R\mathbf{x}+d=\mathbf{x}^T Q\mathbf{x},
\end{equation}
where the matrix $R$ and the constant $\mathbf{d}$ are the result of the multiplication and constant $\mathbf{d}$ can be neglected in the optimisation. Matrix $Q$ now depends on the penalty $\lambda$. The QUBO can then be reduced in a pre-processing step \cite{glover2018logical} to create a smaller problem.

\subsection{Quantum Annealing}
The QUBO is given as input to the QA, which is an important representative of the second paradigm, analogue quantum computing. The premise here is to create an equal superposition over all possible states of a collection of qubits. Then a problem-specific magnetic field, based on the QUBO formulation, is turned on, causing the qubits to interact with each other. Now the qubits move to the lowest energy state, from which the optimal solution of the original problem can be derived. QA has similarities with the well-known Simulated Annealing (SA). However, where SA can only make thermal jumps, QA also uses thermal fluctuations and quantum fluctuations (like tunnelling, entanglement, etc). The most advanced QA is the version of D-Wave Systems (from now called D-Wave). They claim that their devices are practical implementations of adiabatic evolution \cite{farhi2000quantum}. The evolution of a quantum state on the quantum processor of D-Wave is described by a time-dependent Hamiltonian ($H(t)$), consisting of the original Hamiltonian ($H_0$), whose ground state is easy to create, the equal superposition, and the final Hamiltonian ($H_1$), whose ground state encodes the solution of the current problem, via the QUBO. One specific linear annealing schedule is expressed by:
\begin{equation}
H(t)=(1-\frac{t}{T}) H_0+\frac{t}{T} H_1.
\end{equation}
This system is initialised to the ground state of the original Hamiltonian, i.e. $H(0)=H_0$. The adiabatic theorem states that if the system evolves according to the Schrödinger equation, and the minimum spectral aperture of $H(t)$ is not zero, $H(T)$, for $T$ large enough, will converge to the ground state of $H_1$. Although we will not go into the technical details here, it is good to know that it is usually not possible to estimate an adequate time T for which the evolution to the desired state is assured. For some classes of problems, the optimal annealing time has been determined experimentally \cite{albash2018demonstration}. There is therefore no guarantee of optimality. Furthermore, it is not trivial to find a good value for the penalty $\lambda$ and the so-called chain strength; the QA is very sensitive to these parameters. Chain strength is a parameter that results in a penalty if multiple qubits representing one variable do not have the same value in the solution. In addition, D-Wave's state-of-the-art annealer now has 5640 qubits with a connectivity of up to 15. This can model a fully connected clique problem with $n=177$ \cite{mcleod2022benchmarking}. For problems with a lower dependency between the variables, larger problems can be solved. Problems that cannot be placed on a chip of this size will have to be broken into pieces, at the expense of the quality of the solution. D-Wave does offer a standard function that takes care of this decomposition automatically as well as offering hybrid computing pipelines. Note that in the remainder of this paper we use the term Quantum Annealing also if we actually mean the implemented Adiabatic Quantum Computing approach by D-Wave.

\subsection{Quantum Approximate Optimisation Algorithm}
There exists an algorithm that translates QA to the gate-based quantum computer: Quantum Approximate Optimisation Algorithm (QAOA) \cite{farhi2014quantum}, which is a type of variational quantum eigensolver (VQE) algorithm. The VQE is a quantum algorithm that is used often for quantum chemistry, quantum simulations and optimisation problems. It is a hybrid algorithm that uses both classical computers and quantum computers to find the ground state of a given physical system. In QAOA, the adiabatic evolution is approximated by a discretised function that yields an approximation of the ground state of the desired Hamiltonian. The goal of the QAOA algorithm is to find the lowest possible upper bound for the ground state. Here too, we start with an equal superposition over all solutions $\ket{s}$. The algorithm then alternately applies the following two operations, operators, to this state:
\begin{equation}
U(H_0,\beta)=e^{-i \beta H_0} \textrm{ and } U(H_1,\gamma)=e^{-i\gamma H_1},
\end{equation}
where $\beta$ in $[0,2\pi]$ and $\gamma$ in $[0,4\pi]$. If we apply these operators $p$ time, the QAOA produces the following quantum state:
\begin{equation}
\ket{\gamma,\beta}= U(H_0,\beta_p )U(H_1,\gamma_p )… U(H_0,\beta_1 )U(H_1,\gamma_1 )\ket{s}.\end{equation}
Using a QC, the expected value $F_p=\bra{\gamma,\beta}H_1\ket{\gamma,\beta}$ can then be calculated, which gives an upper limit for the ground state. If $p\rightarrow \infty$ and given the right choice of angels $\beta_p$ and $\gamma_p$, this approximation will converge to the optimal solution - the exact ground state of the underlying problem. However, finding the right angles is not trivial and if p must be large for a good approximation, the effectiveness of the algorithm is still an uncertain factor.
The QAOA has more freedom than QA in this regard. Another choice for the so-called mixing operator $U(H_0,\beta)$, is the Quantum Alternating Operator Ansatz \cite{hadfield2019quantum}, which can limit the search space, by, for example, setting hard constraints. The QAOA is an example of a variational or hybrid algorithm. Classical parameters are optimised to create a quantum circuit that solves a problem. This class of algorithms are seen to possess the first practical application to gated quantum computers. After all, it only needs a limited number of qubits and the depth of the circuit (the time factor) is also limited if p is not chosen too large. In addition, variational algorithms are fairly resistant to noise, the limited stability of the qubits.

\begin{figure}
    \centering
    \includegraphics[width=8cm]{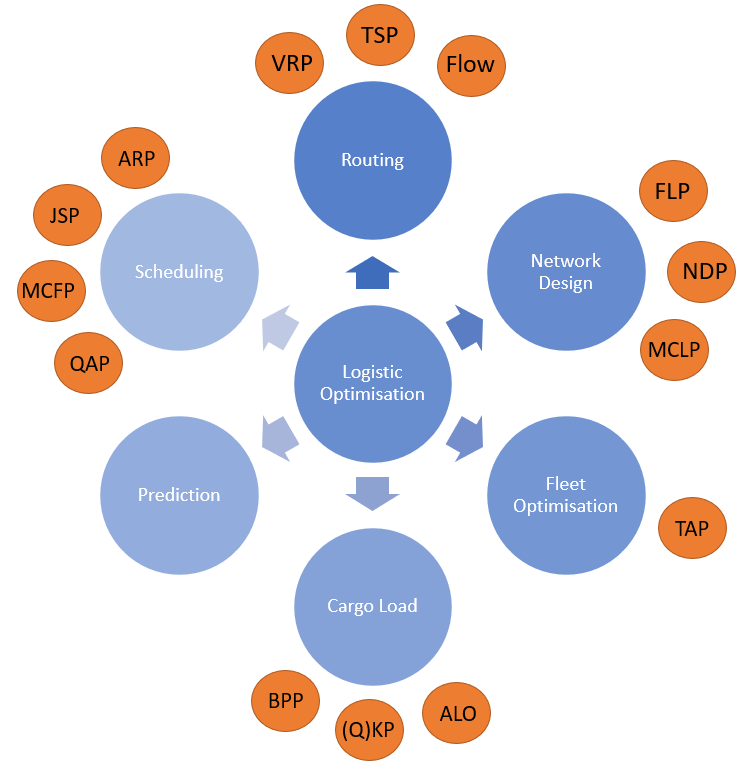}
    \caption{Overview of the six problems areas together with the discussed mathematical problem formulations.}
    \label{fig:overview1}
\end{figure}

\section{Problem overview} \label{sec:overview}
In this section, we delve into the realm of quantum optimisation techniques, focusing on their applications to six pivotal categories of logistics problems: routing, network design, fleet optimisation, cargo loading, prediction and scheduling, see Fig. \ref{fig:overview1}. We will dive into all these categories and discuss the quantum algorithms and approaches suggested for underlying quantitative optimisation problems in literature.

\subsection{Methodology}
\begin{figure}
    \centering
    \includegraphics[width=12cm]{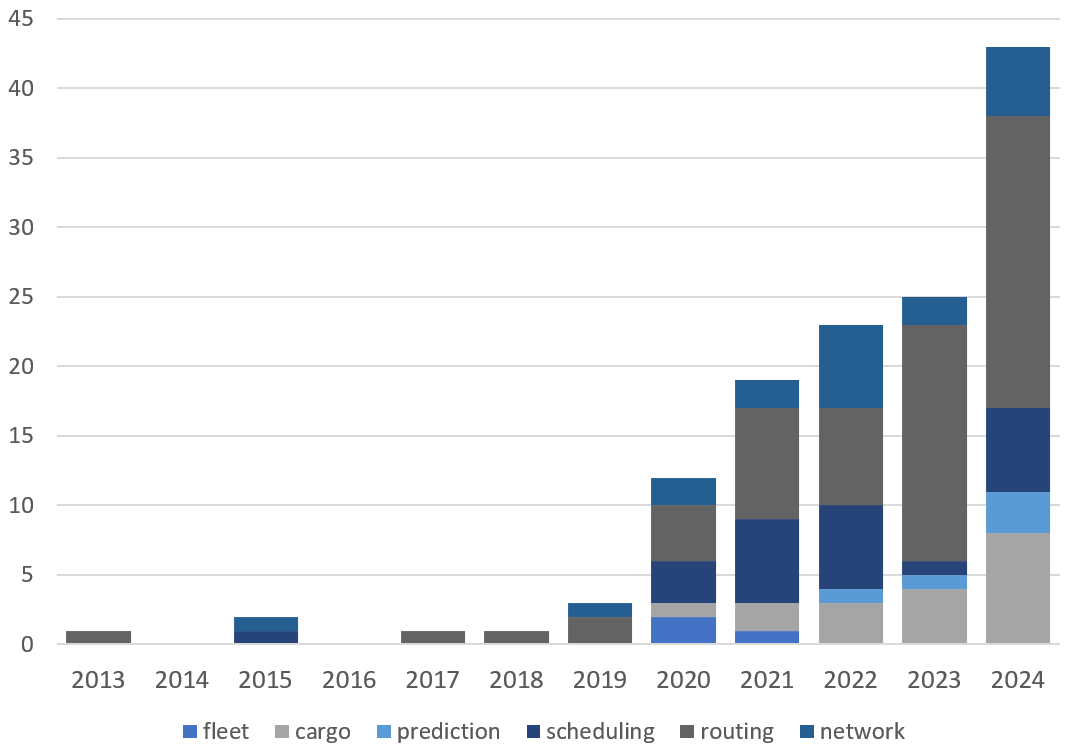}
    \caption{Distribution of the papers over the years, per category: \textbf{fleet} optimisation, \textbf{cargo} loading, \textbf{prediction} and inventory control, \textbf{scheduling}, \textbf{routing} and \textbf{network} design network}
    \label{fig:overview}
\end{figure}

For this work we created an overview that is as complete as possible until 2023. We used Google Scholar and Scopus as source and used the index terms `quantum algorithm' and the category names of the logistic problems. Note that there exists also a lot of literature on `quantum inspired' algorithms like quantum genetic algorithms \cite{xihuai2007application}, quantum evolutionary algorithm \cite{xu2009using} and regular simulated annealing, 
not using the QUBO formulation, such as \cite{chibeles2012simulated,mousavi2013hybrid}. The first QA work we found is from 2013 and 2015, with the works from \cite{chooinvestigating,crispin2013quantum,venturelli2015quantum}. The main flow of research starts around 2020, as shown in Fig. \ref{fig:overview}. 

\subsection{Routing problems}
These problems involve determining the most efficient routes for transporting goods from suppliers to manufacturers and then to customers. It includes the Vehicle Routing Problem (VRP) and the Travelling Salesman Problem (TSP), where you aim to minimise transportation costs while meeting customer demand and satisfying constraints.
\subsubsection{VRP and TSP quantum solutions}
Several studies explore the application of quantum computing to address various aspects of Vehicle Routing Problems (VRP). The following paragraph provides an overview of these works, highlighting their methodologies, findings, and future implications.

One of the earlier works is by Crispin et al. \cite{crispin2013quantum}. They propose an approach to solve the Capacitated Vehicle Routing Problem (CVRP) using a QA algorithm. CVRP is a variant of the VRP where vehicles have limited capacity. The paper discusses the design of a spin encoding scheme for CVRP and an empirical approach for tuning parameters in the QA algorithm. QA is introduced as a metaheuristic that utilises quantum tunnelling in the annealing process. Experiments are conducted on CVRP benchmark instances, and the results are compared with a simulated annealing algorithm. The QA algorithm is found to match or outperform simulated annealing in terms of success rate for various instances. The paper analyses the QA algorithm's performance and discusses the impact of parameters such as population size, temperature, and acceptance ratio.

Srinivasan et al. \cite{srinivasan2018efficient}, introduce a quantum algorithm to solve the TSP. It utilises the quantum phase estimation technique \cite{jiang2021survey}, encoding distances between cities as phases. Unitary operators are constructed based on these phases, and the phase estimation algorithm is applied to estimate eigenvalues representing total distances for all possible routes. Quantum search algorithms are then employed to find the minimum distance and corresponding route, providing a quadratic speedup over classical brute force methods. The algorithm is illustrated with an example of TSP involving four cities, and simulations are conducted using IBM's quantum simulator. 

Liu at al. \cite{liu2024quantum} introduce a hybrid quantum-classical approach for solving the TSP using path-slicing strategies and quantum local search. By dividing the TSP into smaller subproblems and solving them with quantum annealing and  classical solvers, the method efficiently handles computational challenges, achieving near-optimal solutions with improved resource utilization.

The work by Feld et al. \cite{feld2019hybrid} also investigates the CVRP using D-Wave's quantum annealer. They propose a quantum-classical hybrid solution, incorporating a 2-Phase-Heuristic that divides CVRP into clustering and routing phases. Results indicate that the hybrid method can compete with classical construction and 2-phase heuristics in terms of solution quality. However, challenges arise, particularly in finding the `best known solution' for certain datasets, emphasising the need for larger quantum hardware. The study anticipates future improvements in D-Wave's technology, expecting increased qubit connectivity and quantity.

Other hybrid approaches can be found in Mario et al. \cite{mario2024quantum}, presenting a Hybrid Two Step (H2S) and a Hybrid Three Step (H3S) approach and in Sadashivan et al. \cite{sadashivan2024flexible},
that combines quantum annealing for the TSP with classical partitioning. Same holds for Osaba et al. \cite{osaba2024solving} presenting a quantum-classical strategy, Q4RPD, for solving real-world package delivery routing problems.

Borowski et al. \cite{borowski2020new} also delve into (C)VRP, introducing four QA-based algorithms. Their Full QUBO Solver (FQS), Average Partition Solver (AVS), DBSCAN Solver (DBSS), and Solution Partitioning Solver (SPS) showcase promising results. The study employs D-Wave's Leap framework, comparing quantum algorithms with classical metaheuristics. The experiments use established benchmark datasets and custom scenarios with realistic road networks. The authors outline future research directions, including extending quantum formulations to address more realistic variants of VRP.

Weinberg et al. \cite{weinberg2023supply} focus on practical quantum algorithm use, considering the limitations of Noisy Intermediate-Scale Quantum (NISQ) hardware. They adopt a hybrid workflow for solving the CVRP, breaking down problems into smaller instances suitable for QA. Simulated annealing and the D-Wave Hybrid solver are used as quantum substitutes. While not claiming clear advantages over classical methods, they emphasise the sensible application of quantum algorithms to specific bottlenecks, anticipating future advancements.

Kanai et al. \cite{kanai2024annealing} explore a hybrid algorithm combining column generation and annealing to solve combinatorial optimization problems with inequality constraints, specifically the CVRP. The proposed method aims to overcome hardware limitations and achieve better lower bounds through annealing-assisted column generation.

This study by Sinno et al. \cite{sinno2023performance} focus on the performance of commercial QA solvers for the CVRP. The research challenges assumptions made in theoretical studies and simulations on classic hardware, emphasising the need for empirical measurements on real quantum platforms to assess commercial capabilities accurately. The study evaluates the quality of solutions provided by the D-Wave CQM solver for CVRP, considering problem size and complexity. The study suggests that, more than problem size, model complexity significantly affects solution quality, highlighting the importance of minimising constraint density in practical applications. 

Sales et al. \cite{sales2023adiabatic} address the VRP in last-mile logistics, proposing a hybrid solution combining QA and classical techniques. Their analysis, using Amazon Braket and D-Wave's QA, demonstrates the superiority of the hybrid algorithm over full QA, suggesting potential advantages in scenarios with lower data ``clusterability." Despite current challenges, the study identifies a cost-effective relationship between running quantum algorithms and result quality, indicating quantum computing's theoretical advantage in solving constrained clustering problems. \\

The thesis by Palmieri \cite{palmieriquantum} explores the application of quantum optimisation techniques to address the (C)VRP. Here, a heuristic two-phase approach is explored. Nodes are first clustered into disjoint groups, and routing is performed individually within each cluster. Various classical methodologies for the clustering phase are investigated. An alternative approach models the clustering phase as a Modularity Maximization problem, which is tackled using quantum algorithms. The routing phase is modelled as a Travelling Salesman Problem (TSP), and subtour elimination constraints are adaptively added using the Dantzig-Fulkerson-Johnson (DFJ) strategy. This adaptive approach reduces the number of qubits, making the models suitable for both gate-based and annealing-based quantum devices. Various quantum algorithms, such as QA, quantum GAVA (Graver Augmented Multi-seed Algorithm), QAOA and VQE are investigated for quantum optimisation. Gurobi is used as a classical benchmarking method. QA and quantum GAMA outperform the VQE approaches here.

Until here, all works use QA as their solver for the VRP. Rana et al. \cite{rana2022quantum} explore employee transport route and agri-tech logistics optimisation using both a quantum gate-based variational algorithm and QA. Here again, quantum hardware limitations lead to problem simplifications, using problem-specific decomposition techniques, such as  distance-based clustering. The study utilises real-world data and anticipates extending quantum methods to more complex scenarios, acknowledging ongoing improvements in the D-Wave's system for addressing larger, intricate logistics optimisation problems.

Azad et al. \cite{azad2022solving} explore the application of the QAOA for the VRP within Intelligent Transportation Systems (ITS). The paper provides an Ising formulation for the VRP and its variants, adapting them for quantum processing. Despite the limitations of NISQ devices, having 15 qubits, QAOA holds promise for solving combinatorial optimisation problems like VRP. However, they conclude that there is a need for a multi-faceted approach to enhance its performance. Working on aspects such as initialisation techniques, optimising parameter values, exploring different mixer Hamiltonians, and addressing noise-related challenges are essential for realising the full potential of QAOA in practical applications.

The proposed Indirect QAOA (IQAOA) by Bourreau et al. \cite{bourreau2023indirect}, introduces a novel approach, targeting the Hamiltonian of a set of string vectors. It utilises a Quantum Alternating Operator Ansatz with a parameterised family of unitary operators, allowing for efficient modelling of the Hamiltonian. This algorithm addresses the limitations of standard QAOA by incorporating a classical meta-optimisation loop and estimating the average cost of each string vector. The IQAOA approach includes the creation of a quantum circuit with a limited number of gates suitable for current noisy quantum machines. Numerical experiments demonstrate its efficacy in solving the TSP, notably solving 8-customer instances, which are among the largest TSPs solved using a QAOA-based approach.

The study by Mohanty at al. \cite{mohanty2023analysis} presents a VRP solver for three and four cities using VQE on a fixed ansatz. The analysis extends to assess the solution's robustness in noisy quantum channels, considering various noise models. The work contributes to understanding the impact of noise on quantum algorithms for solving real-world optimisation problems. The results highlight the importance of noise analysis for effective use of quantum devices, emphasising the role of optimisers and the need for further experimentation on physical quantum devices with larger VRP instances.

Bentley et al. \cite{bentley2022quantum} also introduce a QAOA based framework for integrating quantum computing with the transport sector. They focus on a CVRP and demonstrate a significant circuit performance enhancement (20 times error reduction) using a real quantum device, emphasising problem decomposition, mapping, and compilation for quantum advantage. Same holds for the approaches in Ramezani et al. \cite{ramezani2024reducing} and Herzog et al. \cite{herzog2024improving}, that propose circuit cutting and other techniques to reduce the circuit size and the number of required qubits.

Next, Xie et al. \cite{xie2023feasibility} also use a QAOA approach for the CVRP. This work explores the application of a variant of QAOA called Grover-Mixer Quantum Alternating Operator Ansatz to solve CVRP. The paper introduces a novel binary encoding for CVRP, focusing on minimising the shortest path while bypassing vehicle capacity constraints. The search space is further constrained using the Grover-Mixer. Experimental results indicate that the proposed method outperforms conventional QAOA in terms of feasibility ratios, optimality ratios, and optimality gaps. Despite promising results, the paper acknowledges drawbacks, such as the requirement for multi-controlled Toffoli gates, which may pose challenges for implementation on near-term quantum devices with noise.

Also Sato et al. \cite{sato2024circuit} use Grover's algorithm for the initial state preparation in quantum circuits, followed by amplification of the optimal solution. Their method leverages Higher-Order Unconstrained Binary Optimization (HOBO) encoding to reduce qubit requirements and improve efficiency.

Also Palackal et al. \cite{palackal2023quantum} look at the CVRP and its relation to the TSP. To facilitate experiments, the CVRP is reduced to a clustering phase and a set of TSPs. The authors extensively test QAOA and VQE on TSP instances, investigating the influence of hyperparameters like classical optimizer choice and constraint penalisation strength. Results suggest that QAOA often falls short in reaching the energy threshold for feasible TSP solutions, even with extensions like recursive and constraint-preserving mixer QAOA. On the contrary, VQE performs better, reaching the energy threshold and showing improved performance. The study evaluates the challenges of quantum-assisted solutions for real-world optimisation problems and proposes perspectives for overcoming these obstacles. It also introduces a performance metric tailored to the problem, focusing on solution quality from an application perspective. 

Li et al. \cite{li2024quantum} explore variational quantum algorithms for collision-avoidance route planning in consumer electronics supply chains. They introduce infeasible solution constraints to optimize qubit usage, integrates vehicle routing with collision-avoidance, and validates the method through quantum simulations, enhancing scalability with a stepwise optimization algorithm.

Last in this category, the paper by Alsaiyari et al. \cite{alsaiyari2023variational} compare two quantum algorithms, general VQE and QAOA, for solving the VRP. The study assesses the maximum VRP instance size that the current state-of-the-art NISQ devices can handle, indicating that large VRP instances exceed current quantum capabilities. While recognising the potential of quantum technology, the study emphasises the need for advancements in both hardware and software for broader adoption.\\

The next paradigm for solving routing problems optimisation is a mixture of quantum computing and machine learning. Sanches et al. \cite{sanches2022short} explore the integration of quantum computing with reinforcement learning, specifically addressing the VRP. Their model introduces quantum circuits in place of classical attention head layers, showcasing competitive performance. While acknowledging it as a proof of concept, the study emphasises the potential symbiosis between quantum computing and machine learning, raising questions about achieving quantum advantage for combinatorial optimisation through (quantum) reinforcement learning agents.

Correll et al. \cite{correll2023quantum} take a different approach, employing quantum neural networks with reinforcement learning for truck routing logistics. The study uses real-world data, the same as Weinberg et al. \cite{weinberg2023supply} we saw earlier, from the automotive sector and achieves results comparable to human decision-making. While the quantum versions of these neural networks offer theoretical speedups, the article acknowledges the need to determine how much of this advantage can be practically realised. Thus, it explores practical realisation methods by exploring methods for tuning quantum circuits and workflow performance.\\

The paper by Xu et. al \cite{xu2024quantum}  proposes a quantum Q-learning model for the CVRP. It uses parameterised quantum circuits to approximate Q-values, enhancing solution quality and efficiency. Experimental results show that the model outperforms classical methods and previous quantum approaches, making it more suitable for NISQ devices.

Also quantum support vector machines are used to solve VRP problems, by Mohanty et al. \cite{mohanty2024solving}. The combine this with variational quantum eigensolvers on 6- and 12-qubit circuits for 3- and 4-city scenarios. Various quantum encoding techniques and optimizers are evaluated using IBM Qiskit simulations.

Lastly we see the rise of quantum variants of the metaphor based meta-heuristics. Qiu et al. \cite{qiu2024novel} introduce a hybrid quantum-classical algorithm that enhances Quantum Ant Colony Optimization (QACO) by integrating a clustering algorithm, specifically K-means. The proposed method was tested using the TSP and demonstrated improved performance across multiple datasets. Additionally, the hybrid algorithm showed robustness to noise, a critical challenge in quantum computing. The study highlights the potential of QACO in the Noisy Intermediate-Scale Quantum (NISQ) era by extending its applicability through classical-quantum integration.

Slightly different and more general versions of the VRP problem are found in the following papers. Ajagekar et al. \cite{ajagekar2020quantum,ajagekar2020quantum2} propose a Hybrid QC-IQFP Parametric Method for solving the Inexact Quadratic Fractional Programming (IQFP) problem associated with the VRP. The hybrid method combines an inexact parametric algorithm and quantum computing techniques, demonstrating efficiency compared to classical Mixed Integer Nonlinear Programming (MINLP) solvers for medium to large-sized VRP instances. The authors highlight the heuristic nature of the method, its potential for improvement with advancements in quantum computing, and its competitive use in solving large-scale IQFP problems.

Masuda et al. \cite{masuda85optimisation} evaluate a quantum-classical hybrid approach for the Time-Dependent Vehicle Routing Problem with Time Windows (TDVRPTW). They employ a QUBO formulation using Fixstars Amplify Annealing Engine (a \textbf{classical}, not quantum, Ising machine). The hybrid approach shows potential efficiency for small-scale problems, emphasising the need for further development to handle larger and more complex instances.

Also Leonidas et al. \cite{leonidasqubit} address the Vehicle Routing Problem with Time Windows (VRPTW). They explore the application of a previously introduced \cite{tan2021qubit} qubit encoding scheme to reduce the number of binary variables. The study utilizes a quantum variational approach on a testbed of VRPTW instances, ranging from 11 to 3964 routes, formulated as QUBO problems based on realistic shipping scenarios. The authors compare their results with standard binary-to-qubit mappings, employing simulators and various quantum hardware platforms, including IBMQ, AWS (Rigetti), and IonQ. The benchmarking is done against the classical solver, Gurobi. Despite the reduction in qubits required, their approach demonstrates the capability to find approximate solutions to the VRPTW comparable to those obtained from quantum algorithms using the full encoding.

Harwood at al. \cite{harwood2021formulating} introduce various mathematical formulations for VRPTW, suitable QAOA, VQE, and alternating direction method of multipliers (ADMM). The formulations are compared from a quantum computing perspective, considering metrics for evaluating the difficulty in solving the underlying QUBO problems. Simulated quantum devices are employed to demonstrate the relative benefits of different algorithms and their robustness in practical scenarios. 

Irie et al. \cite{irie2019quantum} introduce the idea of a time-table, capacity-qubits, and the concept of state into the QUBO formulation. Time-tables are incorporated to represent the temporal aspect of VRP, and capacity-qubits are introduced to manage the capacity of vehicles dynamically. The concept of state allows for the description of different travelling rules based on the state of the vehicles. The paper includes a graphical view of a VRP instance solution obtained using the D-Wave 2000Q. The results demonstrate the effectiveness of the proposed formulation for small-size QUBO systems.

The work by Spyridis et al. \cite{spyridis2023variational}, focuses on the Multiple Travelling Salesman Problem (m-TSP). The study explores the use of the QAOA to efficiently handle the m-TSP, formulating a QUBO problem. The authors highlight potential applications in IoT systems, emphasising the need for innovative error correction techniques to address limitations of NISQ devices. An other variant of TSP is the selective travelling salesman problem (sTSP), which is also known as the orienteering problem (OP), where each destination is associated with a prize. Given a budget that must not be exceeded, the salesperson is not necessarily required to visit all the cities, and his mission is to collect as many prizes as possible. In Le et al. \cite{le2023quantum}, the authors provide a QUBO formulation for this problem and perform experimental results include an analysis of parameters such as annealing time, chain strength, and solution quality for different instances of sTSP with varying numbers of cities.

Salehi et al. \cite{salehi2022unconstrained} and Papalitsas et al. \cite{papalitsas2019qubo} both look at the Travelling Salesman Problem with Time Windows (TSPTW). The TSPTW involves finding a tour with the minimum cost, where each city must be visited within a specified time window. Papalitsas et al. give a first QUBO formulation of this problem. Salehi et al. introduce three different formulations. The first formulation considers both earliest start times and due times for each city. The second formulation presents a higher-order binary model that is more space-efficient. The third formulation is based on an alternative QUBO formulation derived from integer linear programming. These formulations allow for multiple assignments of binary variables to encode the optimal route without penalty, and they can be adapted for other variants of the TSP.
The study by Salehi et al. investigates the efficiency of edge-based and integer linear programming (ILP) formulations through experiments on the D-Wave Advantage quantum annealer. The results indicate that the ILP model outperforms the edge-based model, particularly for instances with four cities, showcasing a higher probability of obtaining samples encoding the optimal route. The paper suggests a natural progression for future work, involving a more nuanced exploration of penalty values for time window constraints and the investigation of alternative techniques for quadratisation in the node-based formulation. 

Marsoit et al. \cite{marsoit2021quantum} address the VRP with uncertain data, proposing a quantum computing approach. In real-world scenarios there are uncertainties, which impact the effectiveness of VRP solutions, think of varying travel times due to traffic congestion, uncertain customer demands, service disruptions, or weather conditions. Failing to account for these uncertainties may lead to inefficient route planning, missed time windows, increased operational costs, and customer 
dissatisfaction. The study formulates the problem, compares it with classical optimisation challenges, and claims to demonstrate the effectiveness of the quantum approach through a numerical example. However, it is not  clear which quantum approach they use. The research acknowledges current limitations of quantum hardware and calls for further advancements.

Dixit et al. \cite{dixit2023quantum2} also discuss VRP with uncertainty, addressing the Stochastic Time Dependent Shortest Path (STDSP) problem. In STDSP, uncertainties arise from factors like demand and supply fluctuations, making the representation of link travel time as a random variable necessary. The study introduces a Quadratic Constrained Binary Optimisation Problem for the STDSP. The research explores the efficiency of QA in comparison to classical solvers, specifically CPLEX. The results indicate that quantum computing, despite its current limitations in terms of qubit numbers and noise susceptibility, exhibits linear computational efficiency.

Fitzek et al. \cite{fitzek2021applying} explore the use of QAOA for the Heterogeneous Vehicle Routing Problem (HVRP). The HVRP involves determining optimal routes for a fleet of vehicles with varying capacities to deliver goods to customers. They formulate the HVRP as an Ising Hamiltonian, showcasing scalability with increasing problem size. The study employs QAOA, which requires defining a cost function and proposing an ansatz for optimization. The study emphasises the trade-off between classical optimiser performance and runtime, providing insights into the potential of quantum computing for solving large instances of HVRP with advanced hardware.

Lo et al. \cite{lo2021genetic} introduce a quantum random number generator (QRNG) in solving pollution-routing problems (PRPs) for sustainable logistics. The hybrid model incorporates a modified k-means algorithm and a genetic algorithm, showcasing the applicability of quantum random number generation in optimising routes for emission reduction. The study suggests future research directions involving metaheuristic algorithms and advancements in quantum computing technology.

The work by Harikrishnakumar \cite{harikrishnakumar2020quantum} introduces a QA approach to address the Multi-Depot Capacitated Vehicle Routing Problem (MDCVRP). The MDCVRP involves assigning routes to vehicles from multiple depots, each with varying capacities, to serve spatially distributed customer locations while satisfying capacity constraints. The authors formulate the MDCVRP and its dynamic version, the D-MDCVRP, as QUBO problems. The authors discuss the complexity of the formulated problems, considering the total number of decision variables. They also present a step-by-step solution framework for solving QUBO formulations on QA hardware. The authors outline future work, which includes solving the QUBO formulations on D-Wave (here the 2000Q), comparing results against classical heuristic algorithms like Tabu Search, and investigating the scalability of the QA approach concerning the number of vehicles and depots. 

The works of Warren \cite{warren2020solving} and Osaba et al.\cite{osaba2022systematic} gives an overview of VRP and TSP implementations.
Warren conducts an analysis of four software programs designed to solve the symmetric TSP, among which D-Wave's solutions and the work by Feld \cite{feld2019hybrid} we saw before. Osaba et al give three main observations in the context of quantum optimisation and routing problems. Regarding the current state of routing problems they indicate that traditional routing problems like the TSP and VRP have been extensively studied, but their formulations often fall short of addressing advanced or realistic scenarios. Researchers face challenges in tailoring problems to hardware capacities, leading to limitations in addressing complex formulations.
The second observation is the current state of the hardware. current commercial quantum devices have inherent limitations such as noise and decoherence. Researchers have worked on error correction and mitigation strategies, but the focus on this aspect is relatively low.
The last observation is on the parameterisation sensitivity in quantum devices. Quantum devices are highly sensitive to parameterisation in various aspects, including problem formulation, algorithm tuning, and hardware specifications. Despite this sensitivity, a majority of practical papers focus on solving specific problems without deep exploration of parameter fine-tuning. Proper parameterisation, especially the choice of (QUBO) penalty values in problem formulation, is highlighted as a significant challenge.

Lastly, the work by Poggel et al. \cite{poggel2023recommending} introduces a framework designed to automate and optimise the decision-making process in quantum-assisted optimisation for real-world problems. Addressing the challenges of formulating, encoding, and selecting algorithms and hardware options, the proposed abstraction layer aims to make quantum-computing-assisted solution techniques accessible to end-users without requiring in-depth knowledge of quantum technologies. The framework incorporates state-of-the-art hybrid algorithms, encoding, and decomposition techniques in a modular manner, allowing evaluation using problem-specific performance metrics. Tools for graphical analysis of variational quantum algorithms are developed, and classical, fault-tolerant quantum, and quantum-inspired methods can be included for a fair comparison. The approach is demonstrated and validated on the CVRP. 

\subsubsection{Other Routing problems}
In addition to addressing the classic Vehicle Routing Problem (VRP) and Travelling Salesman Problem (TSP), quantum computing has been applied to various other routing problems, showcasing its potential in optimising complex logistical scenarios. The first example is the shipment re-routing problem by Yarkoni et al. \cite{yarkoni2021solving}, addressing the challenge of partially filled trucks transporting shipments between a network of hubs. The goal is to minimise the total distance travelled by these trucks, thus reducing fuel consumption and enhancing cost efficiency. The problem instances and optimisation techniques are based on real-world data from an existing shipment network in Europe.
The authors create a QUBO formulation and then employ both classical and hybrid quantum-classical algorithms to solve these QUBOs. The quantum platforms, particularly D-Wave, are found to be promising for addressing this logistics problem due to their qubit support, compared to gate-based and quantum-inspired optimisers.
The article mentions that due to qubit limitations in current quantum hardware, the experiments are conducted with minimal parameters. It anticipates that as quantum hardware improves, more complex real-world problems can be solved using quantum computing. 

Azzaoui et al. \cite{azzaoui2021quantum} extended the scope to smart logistics systems, proposing Quantum Approximate Optimization Algorithm (QAOA) for route optimisation. They emphasise the importance of considering, next to cost and time, also carbon footprint in logistics operations. They propose the deployment of QAOA to calculate and optimise travel paths, selecting the best route to reduce costs, time and carbon footprint, in a \textbf{multi objective approach}, in smart logistics systems.
Next to this approach, they suggest implementing a private blockchain network in smart logistics systems to ensure secure and private communication between different tiers of the smart logistics system, creating a trustworthy cluster. They then conduct simulations, using IBM Quantum Lab and Qiskit, along with blockchain-based secure communication simulations using IBM Hyperledger. The results show promising outcomes, including a 91\% success rate in route optimisation, a 10\% cost reduction, a 6\% reduction in carbon footprint, and improved system scalability. The paper concludes by highlighting the potential benefits of the proposed framework, aiming to improve the scalability, reduce the carbon footprint, and cut down transportation and logistics costs in smart logistics systems. 

The article by Atchade et al. \cite{atchade2021qrobot} introduces quantum computing to the field of robotics, specifically focusing on optimising distance travelled by the robots in warehouses and distribution centres where order picking and batching is performed. The authors develop a real-time quantum optimisation algorithm for this purpose, implemented on a Raspberry Pi 4 as a proof of concept. The proposed system aims to minimise distance travelled and optimise order batching. 
The system allows for hybrid computing, combining quantum and classical processing. The environment developed enables the execution of quantum algorithms on IBMQ, Amazon Braket (D-Wave), and Pennylane, either locally or remotely. The study indicates that Amazon Braket has better time performance than Qiskit or Pennylane. They discuss potential improvements and considerations for applying the system to scenarios with multiple robots and batches.

In the aviation sector, Makhanov et al. \cite{makhanov2023quantum,makhanov2024quantum} dive into the complexities of optimising flight paths, emphasising the use of quantum computing to overcome computational challenges associated with this crucial aerospace engineering operation. The research introduces a customisable modular framework designed to accommodate specific simulation requirements. It also examines the application of quantum-enhanced algorithms using Grover's algorithm, on various quantum architectures and simulations using CPUs and GPUs. The study presents results from running quantum algorithms on IBM hardware, highlighting the potential advantages and challenges associated with quantum computing in the aerospace industry and acknowledging the need for further innovation to practically achieve the theoretical speedup promised by quantum algorithms. Khan et al. \cite{khan2024optimizing} optimise flight routes using a combination of Grover's Algorithm and the Quantum Approximate Optimization Algorithm (QAOA). Grover's Algorithm speeds up the search for optimal routes, while QAOA refines these routes based on real-time data like weather and air traffic. This hybrid approach enhances fuel efficiency, operational effectiveness, and adaptability to dynamic conditions.

The paper by Neukart et al. \cite{neukart2017traffic} focuses on a real-world application, namely traffic flow optimisation in dense road networks. The authors describe the process of mapping a traffic flow optimisation problem onto a QUBO formulation, to demonstrate the feasibility of using quantum computing for real-time, time-critical optimisation tasks, such as continuous redistribution of position data for cars in busy road networks. Due to the limited size and connectivity of current D-Wave's QPUs, a hybrid quantum and classical approach is employed to address the traffic flow problem. The quality of solutions is assessed by counting the number of congested roads after optimisation. The study compares the outcomes against random assignments of routes. The results indicate that the hybrid quantum approach redistributes traffic in a way that reduces the number of congested roads compared to both random assignments and the original route assignment. This work continued by Salloum et al. \cite{salloum2024mini} proposing a method to decompose large traffic problems into smaller sub-problems, leveraging QA for improved speed and effectiveness in managing urban traffic flow, demonstrated through experiments with up to 500 cars. The same authors explore in \cite{salloum2024quantum} optimizing urban traffic flow using Quantum Annealing (QA) to address congestion. Now they employ a hybrid quantum-classical approach, transitioning to a purely quantum method for smaller Quadratic Unconstrained Binary Optimization (QUBO) problems.  

Last example in the routing section, is the work by Phillipson et al. \cite{phillipson2021multimodal}. Also this study presents a real-world application in multimodal container planning and demonstrates how to map this problem to a QUBO formulation for practical implementation on the quantum annealer produced by D-Wave. It describes the implementation of the formulation on both SA and D-Wave's QPU. Due to current limitations in the size of quantum computers, the authors initially formulate the problem with a limited number of decision variables and later propose expansions for larger-scale applications. The conclusion emphasises the need for re-formulating real-world problems to match the QUBO formulation and highlights the potential of quantum computing for solving such problems as the technology matures. The implementation issues on D-Wave's QA, such as finding the right embedding, defining chain strength, and penalty functions, are addressed through a parameter grid search. The paper recommends further research on finding more general methods for these parameters, extending the set with alternative paths, and introducing a quantum variant of column generation for optimising paths.

\subsection{Logistic Network Design}
At the heart of logistical optimisation lies the strategic design of logistic networks, a complex process that involves careful planning, analysis, and decision-making. Logistic Network Design (LND) explores how organisations can craft and refine their supply chain infrastructure to enhance efficiency, reduce costs, and meet the ever-evolving demands of the market. From the selection of distribution centres to the configuration of transportation routes, this exploration aims to unravel the strategic considerations that underpin a resilient and responsive logistical framework. One important classical problem within this area is the Facility Location Problem. For this, Mahasinghe \cite{mahasinghe2021qubo} gives a basic QUBO formulation.

The first work on LND by Ding et al. \cite{ding2021implementation,ding2019logistic} discusses the application of QA to address logistic network design problems (NDPs), which are described as abstract optimisation problems aimed at finding the optimal configuration of supply chain infrastructures and facilities based on customer demand, while minimising costs. They translate the cost function with constraints into an Ising problem and benchmark their results by measuring the accuracy of the solutions against optimal published solutions, with an average error of less than 1\%. Additionally, they compare the performance of their quantum approach with classical algorithms and observe a significant reduction in the number of iterations required. The article emphasises that, even though current quantum annealers are not yet achieving quantum supremacy, they can effectively address relevant supply-chain problems. Following the authors, this demonstrates the potential for state-of-the-art quantum annealers to encode and solve complex logistics problems, despite limitations in quantum hardware. Furthermore, the authors introduce an alternative approach called the combined QA algorithm, which utilises two layers with feedback-control interaction between them. This approach is tested on 12 NDP instances using both quantum annealer simulators and the D-Wave QA, yielding positive results in terms of accuracy and performance compared to classical algorithms. 

The work by Dixit et al. \cite{dixit2023quantum} discusses the Transport Network Design Problem (TNDP), involving optimising network capacities, scheduling maintenance, and identifying new links under resource constraints. Traditionally, meta-heuristic methods like Tabu Search are employed. The paper formulates TNDP as a bi-level problem, with the upper level represented as a QUBO problem, solved using QA on D-Wave's quantum annealer. The results are compared with Tabu Search, demonstrating significant computational benefits using QA.

Next, Malviya et al. \cite{malviya2023logistics} explore the potential of QA for a parcel distribution centre network optimisation using data instances of different sizes. Using a small data instance that fits the current technology, they use D-Wave’s built-in hybrid quantum-classical Kerberos sampler via AWS Braket for solving the problem. An iterative hybrid approach for improving the quality of solutions obtained from the Kerberos sampler is used and then the solutions are compared to the random sampler, a greedy heuristic, an exact solver and SA. Improvement in objective value over random has been shown for different approaches over different data instance sizes. Kerberos sampler performs better than greedy heuristic and at par with exact solver for some data instances. The same authors \cite{malviya2024redesign} presents a quantum approach using QAOA to redesign last mile delivery networks. It formulates the problem with hard constraints, optimizing distribution center operations and customer allocations. Results show improved efficiency and scalability for various problem instances.

Gabbassov et al. \cite{gabbassov2022transit} address the challenge of balancing service efficiency and accessibility in urban transit facility planning. The focus is on transit facility consolidation as a cost-effective strategy to enhance service quality. For this, they propose an optimisation framework that integrates Geographical Information Systems (GIS), decision-making analysis, and quantum technologies. The framework includes a mathematical model capturing non-linear interactions between facilities, demand nodes, inter-facility competition, ridership demand, and spatial coverage, solving the Spatial Interaction Coverage (SIC) problem.
The QUBO based model can, as they suggest, be implemented on various quantum optimisation metaheuristics, including QA, quantum-inspired Digital Annealing, Coherent Ising Machines, and universal GBC.
The framework is applied to the public transit facility redundancy problem in the British Columbia Vancouver metropolitan area and demonstrates a 40\% reduction in the number of facilities while maintaining the same service accessibility, thereby improving efficiency, using a D-Wave Hybrid quantum-classical solver. Additionally, numerical experiments on a synthetically generated dataset demonstrate that the Hybrid solver yields statistically superior demand coverage compared to a SIC problem solved with the APOPT\footnote{Advanced Process OPTimizer} solver.

Also Wang et al. \cite{wang2024variational} solve the FLP. Their paper introduces the Variational Quantum Algorithm-Preserving Feasible Space (VQA-PFS) for solving the Uncapacitated Facility Location Problem (UFLP). VQA-PFS combines mixed operators and Hardware-Efficient Ansatz to enhance success probability and reduce circuit depth, outperforming QAOA, QAOA+, and HEA in efficiency and scalability.

A special case of FLP is the work by Chiscop et al. \cite{chiscop2020hybrid}. This paper presents a hybrid solution method for the Multi-Service Location Set Covering Problem (MSLSCP). It formulates the problem as a Quadratic Unconstrained Binary Optimization (QUBO) problem and uses D-Wave's quantum annealer combined with classical optimization techniques to improve solution accuracy and performance.

A new popular topic here is the charging station location problem (CSLP). Radvand et al. \cite{radvand2024quantum} present a quantum search-based optimization algorithm for solving this problem. By utilizing Grover’s Adaptive Search and Quantum Phase Estimation, the algorithm offers a quadratic improvement in complexity over classical methods, optimizing charging station placement with enhanced efficiency in large transportation networks. Also Sakib et al. \cite{sakib2024quantum} provide a solution strategy for this problem using a QA approach. The model, validated through a real-world case study and solved on D-Wave quantum computers, demonstrates that QA can efficiently identify optimal EVCS locations, improving service quality for users.

The paper by Klar et al. \cite{klar2022quantum} addresses the challenges of factory layouts, emphasising their impact on operational costs. The problem involves assigning functional units to positions, optimising transportation distances. Manual layout planning is time-consuming due to the complexity of considering multiple parameters simultaneously. The study introduces a QA-based approach for factory layout planning, involving the formulation of the problem as a QUBO problem. The results show that QA can solve layout planning problems of different sizes within seconds, overcoming the trade-off between solution quality and computation time.\\

We now shift from network design to location assignment problems (LAP). In the first paper on this theme, Satori et al. \cite{satori2023quantum} propose a novel approach for solving the LAP in an Automated Storage and Retrieval System (AS/RS) using QA. The problem involves optimising the assignment of product locations on shelves to improve picking efficiency. The proposed method considers product pairs based on picking frequency and assigns them to empty shelves in order of distance from an outlet, with the decision variable being the swapping of positions within product pairs. The paper introduces two formulations: the All-Layout formulation and the Swap formulation. The former is formulated as a mixed-integer programming (MIP) problem, while the latter is designed for iterative optimisation, reducing the number of variables. The authors evaluate the performance of both formulations using various solvers, including a MIP-solver (COIN-OR),SA, QA, and hybrid QA. Numerical experiments validate the expected retrieve time as an objective function and demonstrate the effectiveness of iterative optimisation in the Swap formulation. The paper concludes that the proposed Swap formulation, particularly when combined with QA, offers a promising approach for solving this specific problem. 

Next, the chapter by Guo \cite{guo2020quantum} discusses the application of QA to solve, what they call, NP-hard spatial optimisation problems. Spatial optimisation involves mathematical models to find the best solutions for spatial decision problems, such as the FLP and NDP. The case study involves programming a $p-$median model using the D-Wave QA and applying QA to a biomass-to-biofuel supply chain optimisation problem. Numerical results suggest a computational advantage of QA over classical simulated annealing in certain scenarios. The authors suggest that quantum computing, particularly QA, holds promise for solving complex geospatial problems, but further research is needed to understand its full potential and integrate it with modern High Performance Computing HPC architecture. The authors envision a near-term future where QPUs could supplement HPC for specific spatial optimisation problems, with a long-term potential to be a game changer in solving computationally demanding geospatial problems.

The paper by Giraldo et al. \cite{giraldo2022using} contributes to the integration of quantum computing into urban and regional science, focusing on the maximal covering location problem (MCLP). The MCLP involves optimising population coverage within a specified service distance by locating a fixed number of facilities. The paper outlines the process of transforming the MCLP into a format suitable for two quantum computing paradigms: QA (D-Wave) and GBC (IBM). The authors conduct computational experiments on real quantum devices and classical quantum simulators, specifically applying QA and QAOA. The results reveal the successful solution of MCLP instances using QA technology. The paper compares the performance of QA and QAOA, noting that QA shows great potential for solving this kind of problems, while gate-based quantum technology faces (relatively more) challenges.

Facility Location Problems have connections with the Quadratic Assignment Problem. The quadratic nature of this problem makes it, expectedly, suited for quantum computing. The paper by Khumalo \cite{khumalo2022investigation} discusses the application of quantum computing to solve two specific problems: the TSP and the Quadratic Assignment Problem (QAP). The TSP was already discussed before. The authors explore again the potential of VQE and QAOA, in solving instances of TSP and QAP. The paper extends previous benchmarks by including the newest and largest IBM quantum devices. It introduces preliminary findings for the QAOA on tractable instances and presents metrics to assess the feasibility spectrum of quantum algorithms on IBM devices. The authors investigate the impact of the conditional reset feature on IBM systems, comparing it to previous results. The experimental results demonstrate that classical optimisation algorithms outperform quantum algorithms in terms of success rate, feasibility, and computational time for solving TSP and QAP instances for now. The VQE algorithm performs better than QAOA in terms of solution quality metrics. The conditional reset feature on IBM devices does not show a noticeable improvement in computational performance. The paper suggests that as quantum technology evolves, the performance of quantum algorithms could improve. The authors highlight the need for further exploration of quantum formulations, such as QUBO and Alternating Direction Method of Multipliers (ADMM), and the potential advancements in higher-performing processors with more qubits. 

Also Tosun et al. \cite{tosun2022new} dive into the QAP. Their goal is to minimise the total assignment cost in a scenario with $n$ facilities and $n$ locations. This cost is calculated by summing the facility placement costs and multiplying inter-location distances with flow amounts among facilities. The paper discusses Quantum Bridge Analytics, Quantum Computing, and the mathematical formulation of QAP as a QUBO model. It explores the challenges and potential of combining metaheuristics with quantum computing for solving optimisation problems efficiently. The authors propose a methodology to transform large QAP instances into QUBO models and compare results with classical robust tabu search methods. The study suggests that even with current quantum computing limitations, QUBO transformation can be feasible for difficult instances, paving the way for hybrid techniques that combine classical and quantum computing. The authors acknowledge challenges such as finding optimal penalty values and plan to automate this process in future research. 

As a last example, the work of Choo et al. \cite{chooinvestigating} explores the feasibility of solving the FLP as a Quadratic Assignment Problem (QAP) on D-Wave's QA. The authors propose a method to convert the QAP into a QUBO formulation and embed it onto D-Wave's hardware. The study reveals that, in the worst case, the current hardware can only handle QAPs with seven variables due to the limitation of qubits. To address this, the authors demonstrate a reduction in qubits required for QAPs with duplicate rows, potentially allowing the solution of larger instances. Additionally, they introduce a decomposition approach for solving generic QAPs on D-Wave's QA. The work emphasises the challenges and limitations of directly solving QAPs on D-Wave's hardware and provides insights into potential improvements as quantum computing technology evolves.

\subsection{Fleet maintenance and optimisation}
In the field of fleet maintenance and optimisation, the application of quantum algorithms emerges as a promising avenue for addressing complex logistical challenges. Fleet management involves decisions on vehicle maintenance schedules and resource allocation, which are integral to ensuring operational efficiency and cost-effectiveness. The main problem in this area for which quantum approaches are proposed is the Tail Assignment problem (TAP). The Tail Assignment problem (TAP) is faced by airlines, where the objective is to efficiently assign individual aircraft to a set of flights to minimise overall costs. The TAP is part of the larger airline planning process, involving complex optimisation problems with various constraints related to passengers, crew, aircraft, maintenance, and ground staff. 

The first work in this direction is the paper by Vikstaal et al. \cite{vikstaal2020applying}. They explore the application of the QAOA to instances of this problem derived from real-world data. The instances are reduced to fit on quantum devices with 8, 15, and 25 qubits, leaving only one feasible solution per instance. The reduction allows mapping the Tail Assignment problem onto the Exact Cover problem. The QAOA is simulated on an ideal quantum computer to investigate its performance in solving the simplified Tail Assignment problem. The study reveals that repeated runs of QAOA can identify feasible solutions with close to unit probability for all instances. The authors also observe patterns in the variational parameters, enabling the use of an interpolation strategy that simplifies the classical optimisation part of the QAOA. The study also highlights non-trivial properties in the connectivity of the instances and raises questions about the performance of QAOA compared to classical algorithms for larger instances of the problem. The authors also mention an alternative method for optimising variational parameters using the Gibbs objective function, leaving it for further study.

Next, the dissertation of Martins \cite{martins2020applying} aims to study the feasibility of solving the TAP, considering operational restrictions and costs, using QA. The main contributions of the dissertation include detailing a model for the Tail Assignment Problem, analysing and comparing the scalability of different modelling techniques, and presenting a comparison of solutions obtained through classical and hybrid approaches. The scalability analysis indicates that Direct QUBO modelling performs best, and the comparison between solvers suggests that Hybrid Solvers outperform others in terms of solution quality. 

This paper by Willsch et al. \cite{willsch2022benchmarking} conducts a benchmark of the QPUs of two prominent quantum annealers: the D-Wave Advantage (with 5000+ qubits) and its precursor D-Wave 2000Q (with 2000+ qubits). One of the benchmarking problems used is the TAP. The benchmark set encompasses problems of diverse sizes, connectivity levels, and complexities. Results of the study are that the Advantage system exhibits superior performance across almost all problems, showcasing a noteworthy increase in success rates and the ability to handle larger problems compared to D-Wave 2000Q.
The improved performance of Advantage is attributed not only to the increased qubit count but also to enhanced qubit connectivity. Advantage's Pegasus topology, featuring 15 connections per qubit, surpasses D-Wave 2000Q's Chimera topology with only 6 connections. The Advantage system demonstrates approximately twice the speed of D-Wave 2000Q in terms of programming and readout times. It also displays shorter time-to-solution and annealing times, along with consistently higher success rates and smaller fluctuations over multiple repetitions. D-Wave 2000Q may achieve better success rates for problems with sparse connectivity that do not necessitate the many new couplers present on Advantage. Improved connectivity does not universally enhance performance, depending on the specific requirements of each problem instance.

\subsection{Cargo-loading, knapsack and bin-packing problems}
Efficient cargo loading is crucial for industries grappling with space constraints and diverse load requirements. The knapsack problem involves selecting the most valuable combination of items to fit a limited capacity, while bin-packing focuses on optimising the allocation of items into containers. Also here, classical algorithms encounter difficulties in swiftly solving these problems, especially as the scale and complexity increase. This section explores the application of quantum algorithms to approximately solve cargo loading, knapsack, and bin-packing problems.
 
The paper by Nayak et al. \cite{nayak2022quantum} focuses on addressing challenges in the aviation industry related to the optimisation of aircraft cargo loading and the reduction of loading/unloading operations at multiple stopovers. The primary goals are to maximise the payload capacity of aircraft, leading to reduced fuel consumption and increased revenue, and to minimise the operational and handling costs associated with loading and unloading cargo. The paper presents a quantum approach for tackling these challenges, experimenting with both QA-based and GBC algorithms. Among the quantum platforms tested, D-Wave's QA is identified as the most promising, as it offers sufficient qubits to address realistic problems compared to gate-based and quantum-inspired optimisers. The experiments are conducted on a Boeing 747-8F aircraft, one of the largest cargo carriers in the world. The model used for the experiments manages the 2D/3D stacking of containers with and without racks. Results obtained from quantum solvers, particularly D-Wave, are said to outperform classical solvers, demonstrating the potential of quantum computing in solving complex cargo loading problems. The authors emphasise that due to the limitations of qubits in current quantum hardware, their experiments had to be constrained to minimal parameters. 

Also, Sotelo et al. \cite{hernandez2020application,sotelo2021determination} created an application of quantum optimisation techniques to cargo logistics on ships and airplanes. Their papers address the optimisation challenge of determining the optimal configuration for maximising the loading of cargo vehicles while adhering to specific constraints: maximum volume available and maximum possible weight. The proposed solution, presented at the Airbus Quantum Computing Challenge, employs the VQE algorithm. The Airbus Quantum Computing Challenge consisted of 5 problems: 1. Aircraft Climb Optimisation, 2. Computational Fluid Dynamics, 3. Quantum Neural Networks for Solving Partial Differential Equations, 4. Wingbox Design Optimisation  and last 5. Aircraft Loading Optimisation (ALO).
The solution  was implemented in the IBM Qiskit platform, utilising a noise-free environment simulator. The problem set comprises 200 different container arrangements for loading, with random starting points and 1000 iterations for the VQE algorithm. Realistic weight distributions are ensured by selecting weights with an expected value of 1/3 of the maximum weight of the problem instance. Results are evaluated using a metric based on the logarithm of the median of normalised energies for 200 instances. The experiments compare the performance of seven classical optimisers available in Qiskit. The findings indicate variations in the convergence speed of classical algorithms, with the Nelder-Mead algorithm demonstrating superior results. The convergence analysis reveals that over 70\% of problem instances achieve a convergence ratio of 0.01 for the Nelder-Mead algorithm, while other classical algorithms exhibit varying degrees of convergence efficiency. For instance, the COBYLA algorithm achieves a 50\% convergence ratio.  

Also the work of Traversa \cite{traversa2019aircraft} and Pilon et al. \cite{pilon2021aircraft} published a solution to the ALO problem as formulated by Airbus. Unlike developing algorithms for future quantum computers, the approach by Traversa formulates the ALO problem using Integer Programming (IP) and explores its solution through the MemComputing paradigm (which is not quantum). The ALO problem involves optimising the placement of containers with different sizes and weights in an aircraft, considering constraints on maximum weight, shear, and centre of gravity. The study proposes a two-objective IP problem with binary variables, utilising the MemCPU software for efficient solution strategies. The MemComputing approach, based on self-organising algebraic gates, proves to be a non-algorithmic solution with promising scaling properties, demonstrating its potential for solving large-scale ALO problems efficiently.
Pilon formulates this problem using a QUBO formulation. The model's performance is benchmarked on different solvers, including a classical solver for QUBO functions and the D-Wave 2000Q Quantum Annealer. The results indicate the feasibility of the QUBO formulation and highlight comparisons between classical and quantum approaches in terms of computational time and loaded weight. 

There are several papers looking directly at the Bin Packing Problem (BPP). The work of Romero et al. \cite{v2023hybrid} introduces a hybrid quantum-classical framework, called Q4RealBPP, designed to solve real-world instances of the three-dimensional Bin Packing Problem (3dBPP), involving efficient packing of items into bins. The work in \cite{romero2023solving} adds to that framework in the following way: i) the existence of heterogeneous bins, ii) the extension of the framework to solve not only three-dimensional, but also one- and two-dimensional instances of the problem, iii) requirements for item-bin associations, and iv) delivery priorities. Q4RealBPP addresses realistic characteristics such as package and bin dimensions, overweight restrictions, affinities among item categories, and preferences for item ordering, making it applicable to actual industrial and logistics scenarios. They successfully apply Q4RealBPP to 12 instances of different nature, showcasing its capacity to handle real-world constraints, based on (D-Wave's) Leap Constrained Quadratic Model (CQM). They outline the need for a thorough comparison between Q4RealBPP and traditional artificial intelligence methods, considering factors such as robustness of results and execution times.

Another work on the BPP is by De Andoin et al. \cite{de2022comparative,de2022hybrid}. These papers explore the potential of quantum and hybrid quantum-classical algorithms to provide advantageous solutions for the one-dimensional BPP (1dBPP). The proposed hybrid approach integrates a QA subroutine for sampling feasible solutions, followed by a classical optimisation subroutine for problem solution construction. To assess its performance, the paper compares this hybrid strategy with classical alternatives, specifically random sampling and a random-walk-based heuristic. The focus is on evaluating the quantum approach's ability to avoid stagnation, a common issue in classical algorithms. The benchmark comprises 18 instances with 10 and 12 packages, incorporating various weight distributions. Results indicate that the QA strategy avoids stagnation observed in classical algorithms, enabling a faster full sampling of feasible partial solutions. The hybrid approach outperforms the random walk in a majority of cases. To measure the performance of the sampling strategy, a metric counting the number of iterations needed to measure the full feasible partial solution space is proposed. The paper emphasises the importance of hyperparameter optimisation, acknowledging that a more refined selection could further enhance algorithm performance.

The study bu Cellini et al \cite{cellini2023qal,cellini2024qal} introduces QAL-BP, a novel QUBO approach specifically for BPP. QAL-BP stands for Quantum Augmented Lagrangian approach for Bin Packing problem. It utilises an augmented Lagrangian method to incorporate the bin packing constraints into the objective function while also facilitating an analytical estimation of heuristic, but empirically robust, penalty multipliers. This approach leads to a more versatile and generalisable model that eliminates the need for empirically calculating instance-dependent Lagrangian coefficients, a requirement commonly encountered in alternative QUBO formulations for similar problems. To assess the effectiveness of the proposed approach, they conduct experiments on a set of bin-packing instances using D-Wave's Advantage system. Additionally, they compare the results with those obtained from two different classical solvers, namely simulated annealing and Gurobi. The experimental findings confirm the correctness of the proposed formulation and also demonstrate the potential of quantum computation in effectively solving the bin-packing problem, particularly as more reliable quantum technology becomes available. 

Also Gatti et al. \cite{gatti2024qubo} use a BPP formulation to tackle aircraft load optimization. Using QA, they maximizes load characteristics like weight and volume while ensuring flight stability. Results from D-Wave simulations show promising improvements in efficiency and scalability for real-world cargo aircraft models.

Last, Matt et al. \cite{matt2024heuristic} introduce a QAOA approach for solving a BPP-alike problem, the irregular strip packing problem, optimizing material usage in industries. The approach decomposes the problem into two sub-problems: the TSP and rectangular packing. Using quantum optimization algorithms like QAOA and QOBA, the algorithm minimizes waste and improves efficiency, outperforming classical methods in experiments.

The next related problem is the Knapsack Problem (KP). Van Dam et al. \cite{van2021quantum} introduces enhancements to the QAOA, to make it more suitable for Knapsack Problem (KP). The KP involves selecting items with weights and values to maximise the total value under a weight constraint. The contributions of the paper are two techniques applied to QAOA. They first use the outcome of a classical greedy algorithm to define an initial quantum state and mixing operation for QAOA. Next they use quantum exploration to avoid local minima around the greedy solutions obtained from the classical algorithm. The study compares the performance of constant-depth quantum optimisation heuristics with similarly shallow classical greedy and simulated annealing algorithms. Despite the likelihood that more sophisticated classical algorithms may outperform quantum algorithms, the comparison is considered fair as it involves algorithms of equal simplicity. The paper suggests that the quantum algorithms perform well with weak dependency on instance-specific fine-tuning. Christiansen et al. \cite{christiansen2024quantum} integrate the Quantum Tree Generator (QTG) with a Grover-mixer QAOA framework and present a new method called Amplitude Amplification-mixer QAOA (AAM-QAOA) for solving the KP. Experimental results show that AAM-QAOA outperforms the state-of-the-art Copula-QAOA for smaller instances, but both methods struggle for larger instances at lower circuit depths. However, for sufficiently high circuit depths, AAM-QAOA can eventually sample the optimal solution, offering promising results for optimizing combinatorial problems like the knapsack problem.

The same problem, however in a slightly different application area, is studied in the work by Benson et al. \cite{benson2023cqm}. This paper demonstrates the use of D-Wave's Leap Hybrid solver for solving a KP, specifically selecting meal combinations from a fixed menu that fit certain constraints. The optimisation problem is formulated as a Constrained Quadratic Model and uses D-Wave's CQM solver. The paper also discusses the generalisation of this model for finding optimal drug molecules within a large search space with complex and contradictory structures and property constraints. The CQM-solver is shown to solve multi-object optimisation problems on an expedited timescale, making it a valid choice for molecular drug design. While the presented problem has a manageable number of combinations, the authors discuss the scalability of QA for larger optimisation problems, highlighting potential advantages over classical computing. Also Bozejko et al. \cite{bozejko2024optimal} use D-Wave's quantum machine and compute lower and upper bounds, enhancing efficiency and solution quality for KP problems in complex production systems. 

A variation of the KP that is even more likely to benefit from quantum computing is the Quadratic Knapsack Problem (QKP). Bontekoe et al. \cite{bontekoe2023translating} recognise that, in general, constrained quadratic binary optimisation problems can be translated into QUBOs in several ways, which can have a large impact on the performance when solving the QUBO. They show six different QUBO formulations for the QKP and compare their performance using simulated annealing. The best performance is obtained by a formulation that uses no auxiliary variables for modelling the inequality constraints.

Different approaches can be found the next works. In Shirai et al. \cite{shirai2024post} a postprocessing variationally scheduled quantum algorithm (pVSQA) for constrained combinatorial optimization problems is proposed. It combines variational methods and postprocessing techniques to improve solution quality and efficiency. Applied to graph partitioning and quadratic knapsack problems, pVSQA demonstrates superior performance on both simulators and real quantum devices.
Ardelean et al. \cite{ardelean2024hybrid} introduce a hybrid quantum search algorithm with genetic optimization (HQAGO) to solve complex KPs efficiently. By fixing some qubits as classical bits, it reduces the search space and computational complexity.
Lastly, Cui at al. \cite{cui2024hybrid} propose a hybrid quantum search algorithm for the Multi-Dimensional Knapsack Problem (MDKP). It combines Grover's search with classical preprocessing and iterative refinement, reducing depth, width, and runtime. The method shows improved efficiency and scalability compared to existing quantum and classical algorithms.

\subsection{Prediction and Inventory Control}
The next category of application is that of demand prediction. There are many papers on Quantum Machine Learning (QML) and prediction, however only one on a direct application within logistics and supply chain optimisation. Jahin et al. \cite{jahin2023qamplifynet} focus on improving backorder prediction in supply chain management (SCM) through the introduction of a novel hybrid quantum-classical neural network called QAmplifyNet. Backorder prediction is used for optimising inventory control, reducing costs, and enhancing customer satisfaction. Traditional machine-learning models face challenges with large-scale datasets and complex relationships. The proposed QAmplifyNet demonstrates high accuracy in predicting backorders, particularly on short and imbalanced datasets, outperforming classical models and quantum neural networks.
The study also addresses the scarcity of research on product backordering, emphasising the challenges of class imbalance in datasets. While classical machine learning models have been widely used, following the authors, quantum-inspired techniques show superior performance, specifically QAmplifyNet. The research integrates quantum and classical computing to leverage the strengths of both, overcoming challenges in handling short, imbalanced datasets common in SCM.
The methodological framework involves evaluating seven preprocessing techniques, selecting the best-performing one, and enhancing the model's interpretability using Explainable Artificial Intelligence (XAI) techniques. Practical implications include improved inventory control, reduced backorders, and enhanced operational efficiency. QAmplifyNet achieves a high F1-score of 94\% for predicting ``Not Backorder" and 75\% for predicting ``Backorder", along with the highest AUC-ROC score of 79.85\%. The hybrid quantum-classical model shows promise for real-world SCM applications, paving the way for further exploration of quantum-inspired machine learning in supply chain management. 

The work by Jiang et al. \cite{jiang2022quantum} highlights the challenges in classic computing related to large state and action spaces in supply chain problems. They introduce a quantised policy iteration algorithm designed for inventory control problems and demonstrate its effectiveness. They perform simulations and experiments using IBM Qiskit and the qBraid system. They underscore the practicality of variational algorithms for solving small-sized inventory control problems.

A recent book provides some new insides. The chapter by Sehrawat \cite{sehrawat2024predicting} explores the application of quantum machine learning (QML) for demand prediction in supply chain networks. Traditional demand forecasting methods often struggle with the complexities and uncertainties of modern supply chains. QML, leveraging quantum computing’s computational power and the adaptability of machine learning algorithms, offers significant advantages over classical methods. By applying QML techniques, organizations can improve demand prediction accuracy, optimize inventory, and enhance overall supply chain performance. Through case studies and practical examples, the chapter demonstrates how QML can support informed decision-making and drive improvements in supply chain management.
The chapter by Gutta et al. \cite{gutta2024ai} explores the integration of artificial intelligence (AI) and quantum machine learning (QML) to enhance supply chain forecasting. The combination of AI's ability to analyse large datasets and extract insights with QML's capacity for processing complex probabilistic distributions offers unparalleled accuracy in demand forecasting, inventory optimization, and risk mitigation. By leveraging these technologies, organizations can transform their forecasting processes, improve operational efficiency, and enhance competitiveness. The chapter discusses the synergistic potential of AI-infused QML models, demonstrating their impact through case studies and real-world applications in supply chain management.
Last, the chapter by Koushik et al. \cite{koushik2024enhancing} investigates the application of deep learning strategies to advance predictive maintenance within supply chain systems. Predictive maintenance plays a critical role in ensuring the reliability and efficiency of equipment and machinery across supply chain operations. Traditional approaches often struggle to effectively capture complex patterns and anticipate impending failures. However, deep learning techniques offer a promising solution by leveraging neural networks to analyze vast amounts of sensor data and identify early indicators of potential equipment malfunctions. This chapter explores various deep learning architectures, such as convolutional neural networks (CNNs) and recurrent neural networks (RNNs), tailored for predictive maintenance applications. Through case studies and real-world examples, the chapter illustrates how deep learning strategies can empower organizations to proactively manage asset health, minimize downtime, and optimize maintenance schedules, thereby enhancing overall supply chain resilience and performance.

\subsection{Scheduling}
In this section, we delve into the application of quantum algorithms to address scheduling problems within the domains of logistics and supply chain optimisation. Scheduling plays an important role in orchestrating the efficient flow of resources, managing timelines, and optimising overall operational processes. The research by Riandari et al. \cite{riandari2021quantum} explores the potential of quantum computing in production planning or scheduling, using quantum algorithms like QAOA and VQE to minimise production costs and address demand and resource constraints. They evaluate the potential benefits, limitations, and difficulties associated with employing quantum computing to production planning.

The work by Ajagekar et al. \cite{ajagekar2020quantum,ajagekar2020quantum2} discusses multiple optimisation problems, including the job-shop scheduling problem (JSP). They propose a hybrid approach to address this complex problem, which involves determining an optimal schedule for a set of jobs on a set of machines. The proposed MILP model incorporates due dates and sequence-independent processing times, with decision variables representing start times, assignments, and sequencing of jobs. The hybrid QC-MILP decomposition method involves two phases: solving the MILP problem with a classical Gurobi solver and employing QC for sequencing. The quantum processor, QA, is utilised to determine feasible schedules, overcoming memory limitations of classical solvers for larger instances. Computational results indicate competitive performance, particularly for large-scale industrial problems, where classical solvers face limitations. The approach says to guarantee a global optimal solution but may require longer runtimes, showcasing the complementary strengths of MILP and QC in tackling intricate scheduling problems effectively. The other problem they study is the Manufacturing Cell Formation (MCFP) problem. Cellular manufacturing involves organising equipment to facilitate continuous flow production, enhancing efficiency. Here, they propose a hybrid QC and Mixed-Integer Quadratic Programming (MIQP) approach to solve the MCFP. The MIQP model is formulated to minimise costs associated with intracellular movement, resource utilisation, and machine set-ups. The dual LP model is solved classically, and its results guide the quantum step, solving a QUBO problem. The hybrid method efficiently handles large instances, outperforming conventional solvers. The proposed approach provides optimal or near-optimal solutions, demonstrating again the complementary strengths of classical and quantum computing in tackling complex manufacturing problems. 

The article by Permin et al. \cite{permin2022simple} explores the application of quantum computers to simplify the complexity of Job Shop Scheduling (JSP). The study employs Grover's algorithm in IBM Qiskit to translate a JSP into a binary qubit format, effectively reducing the search space. By conducting simulations, the research demonstrates a significant decrease in complexity

The article  by Venturelli et al. \cite{venturelli2015quantum} also focuses on the JSP. Here, QA is employed as a solver for JSP, with a detailed presentation of the formulation and strategies used. The study discusses the challenges associated with QA, including limitations in precision, connectivity, and the number of variables. The quantum annealer's performance is compared with classical global-optimum solvers, particularly, as the authors call them, algorithms B (based on Brucker et al. \cite{brucker1994branch}) and MS (based on Martin et al. \cite{martin1996new}). Results indicate that D-Wave 2 annealer faces challenges in outperforming classical algorithms on small instances of JSP, raising questions about its scalability and asymptotic advantage. Motivated by these QA results, the study proposes a QA JSP solver, addressing issues such as embedding, pre-processing, and running strategies. The article introduces a solving process for a full JSP optimisation solver, inclusive adapting classical algorithms as pre-processing techniques. The proposed timespan discrimination is highlighted as a flexible compromise between full optimisation and decision formulations, allowing for immediate benefits from precision improvements. 

Also Kurowski et al. \cite{kurowski2020hybrid} employ a QA approach for the JSP.
The mapping of the JSP to the QUBO formulation is detailed, considering the constraints and the specific topology of the D-Wave 2000Q. The proposed heuristic is explained, involving the decomposition of the JSP into smaller optimisation problems and the tuning of QUBO parameters. Experimental studies on a JSP test instance (ft06) are presented, including estimates of the number of feasible solutions. The results obtained from the QA heuristic are compared with optimal solutions generated by classical heuristic methods for the JSP benchmark. 

The study by by Amaro \cite{amaro2022case} explores the application of VQAs on IBM's superconducting quantum processors to solve the JSP. The four VQAs tested are the QAOA, general VQE, variational quantum imaginary time evolution (VarQITE), and the recently introduced filtering VQE(F-VQE). The study compares convergence speed, scalability, and accuracy of the VQAs, with a focus on practical applicability to non-fault-tolerant hardware. The results show that F-VQE outperformed other algorithms in terms of convergence speed and frequency of sampling the optimal solution, even on instances with five qubits. VarQITE exhibited slower convergence but sampled optimal solutions more frequently. VQE converged slowly and sampled optimal solutions less frequently, while QAOA struggled with convergence due to the complexity of circuits and optimiser choice. In a subsequent set of experiments, F-VQE was further tested on larger instances with up to 23 qubits, showcasing its promising performance and ability to handle larger combinatorial optimisation problems. The study suggests that F-VQE is a step towards addressing challenges associated with quantum hardware limitations, including sparse connectivity and cross-talk noise.

Schworm et al. \cite{schworm2023solving,schworm2024evaluation} explore the application of QA as a metaheuristic for solving the JSP, particularly the flexible job shop scheduling problems (FJSP). QA utilises quantum mechanical effects to evaluate multiple solutions simultaneously, potentially providing time-efficient solutions for complex assignment problems. The study assesses QA's efficiency, scalability, solution quality, and computing time by solving FJSP instances of various sizes. The results demonstrate QA's potential in finding high-quality solutions for the FJSP within seconds, even for large instances, making it a promising approach for industry-scale applications. The evaluation includes a scientific benchmark comparing QA with state-of-the-art algorithms, highlighting QA's efficiency in finding high-quality solutions in a short time. While QA shows promise, challenges such as dynamic factors in industrial settings and the need for multi-objective optimisation are acknowledged for future research. The study concludes that QA can be a valuable tool for intelligent transportation systems, emphasising the need for further improvement and exploration of QA's capabilities.

The same authors \cite{schworm2024multi} present a Quantum Annealing-based Solving Algorithm for multi-objective flexible job shop scheduling. It optimizes makespan, total workload, and job priority simultaneously. The method decomposes large problems into subproblems, uses bottleneck factors for job selection, and demonstrates superior performance compared to classical algorithms in terms of solution quality and computation time.

The paper by Denkena et al. \cite{denkena2021quantum} introduces a new method for process-parallel Flexible Job Shop Scheduling (FJSP) using QA. The FJSP requires continuous and process-parallel optimisation of machine allocation and processing sequences. The proposed QA-based method demonstrates results comparable to classical heuristics with reliably low anneal times, especially for large problem instances. The approach is shown to be robust against stochastic influences and hyper-parameter variations. In a practical use case, the approach outperforms classical MES solutions and combinations of MES with genetic algorithms in terms of makespan reduction. 

This paper by Carugno et al. \cite{carugno2022evaluating} also investigates the application of QA to the JSP. The study provides a detailed analysis of applying QA to JSP, covering problem formulation, quantum annealer fine-tuning, and comparison with classical solvers. The research addresses often-overlooked aspects, such as computational costs, qubit requirements, and mitigating chain breaks. Additionally, advanced tools like reverse annealing are explored. Results reveal challenges at various stages, highlighting important research questions and areas for improvement in QA technology. The study contributes to a balanced discussion on QA's potential and limitations in solving JSP, emphasising open research directions for further exploration.

Windmann et al. \cite{windmann2024quantum} introduce a QAOA based approach to job scheduling for automated storage and retrieval systems. It formulates the problem as an asymmetric QUBO and uses QAOA for solutions. Evaluations show promising results in minimising transport costs and improving scalability.

A factory scheduling problem was presented by Yarkoni et al. \cite{yarkoni2021multi}, derived from a real-world optimisation problem involving re-ordering colour sequences in a paint shop during the manufacturing of cars. The task is to assign colours to sequences of cars such that all customer orders are fulfilled while simultaneously minimising the number of colour switches within the sequence. In this work, the authors investigated problem instances ranging from 10–3000 cars in a single sequence. The multi-car paint shop problem described in the paper has a simple Ising model representation, and instances derived directly from real-world data were solved using D-Wave's QPUs, hybrid algorithms, and classical algorithms. The results showed that while the quantum hardware and hybrid algorithms were able to provide adequate solutions to the problem for smaller sizes, the performance of these algorithms approached that of a simple greedy algorithm in the large size limit. However, the analysis also concluded that QA is currently approaching the limit of industrially-relevant problem scales.
Partly the same authors look at the same problem but then using the QAOA approach in \cite{streif2021beating}. This research combines numerical simulations with experimental data obtained from a trapped ion quantum computer, presenting a comparison between the performance of QAOA and classical heuristics, particularly in the infinite-size limit for noiseless quantum computation. The experimental findings show the quick degradation of the quantum algorithm's efficiency as the problem size increases. 

Also Huang et al. \cite{huang2022paint} look at the paint shop problem. While previous research primarily targeted minimising color changeovers between vehicles, this study introduces a novel problem formulation considering not only color transitions but also the impact on product quality. By incorporating a machine learning model, which is not explained in the paper, to predict quality defects and leveraging quantum computing, the paper presents an approach for solving the generalised paint shop sequencing problem. The paper concludes with a case study showcasing the potential of QAOA in solving a small-scale sequencing problem, paving the way for future exploration in real-world, industrial-sized scenarios.

The next problem coming from the aircraft industry, is the Aircraft Recovery Problem (ARP), a crucial aspect of airline disruption management. Given the financial pressures on airlines, aggravated by the recent pandemic, a swift and efficient solution to recover from disruptions is essential to minimise financial losses. Mori et al. \cite{mori2022replanning} develop a Two-Stage algorithm to model ARP as a QUBO problem. Two experiments are conducted to evaluate the model's performance using QC in comparison to classical solvers. The first experiment compares the proposed approach with a previous one \cite{de2015evolutionary} based on evolutionary computation algorithms, showing similar execution times and results. The second experiment compares a classical solver with a hybrid solver using the proposed model, demonstrating identical solutions but with faster execution times for the hybrid solver. 

The thesis by Grange \cite{grange2024design} focuses on designing and applying quantum algorithms to solve complex railway optimization problems. It explores both exact and heuristic quantum algorithms, including Variational Quantum Algorithms and Quantum Approximate Optimization Algorithms, to improve solution quality and computation time for combinatorial optimization challenges faced by railroad companies.

A more generic problem that also has applications in the aircraft industry is the Quadratic Assignment Problem (QAP). Mohammadbagherpoor \cite{mohammadbagherpoor2021exploring} explores the QAP in the application of airline gate-scheduling . Here, the QAP involves optimising the allocation of facilities (airline gates) to locations (flights) with known assignment costs. 
The paper introduces the concept of using VQE and space-efficient graph colouring, to enhance quantum computing algorithms for solving QAPs. The study tests these enhanced quantum computing algorithms on an 8-airline gate and 24-flight test case using both the IBM quantum computing simulator and a 27-qubit superconducting Transmon IBM quantum computing hardware platform. The results suggest that smaller circuit size and efficient mapping to hardware topology enable the application of variational quantum algorithms to larger problem sets, a critical advancement in solving optimisation problems.

The work by Scherer et al. \cite{scherer2021oncall} discusses the application of Grover's algorithm, to solve the ``on-call spacecraft operator scheduling" problem at the German Space Operating Center (GSOC). The problem involves scheduling operators for spacecraft missions, considering various constraints and optimisation goals. The scheduling must allocate approximately 50 operators to 20 positions over 180 days, while satisfying constraints such as operator availability, working preferences, and limiting consecutive working days. Evaluation is done using Qiskit's QASM simulator with up to 32 qubits. The success rates are evaluated for different problem sizes and configurations. The paper successfully demonstrates solving the scheduling problem using Grover's algorithm on a quantum simulator. Future work includes testing on real quantum devices, improving qubit stability, and exploring quantum-classical hybrid approaches. Emphasis on minimising the number of qubits and optimising the algorithm's width and depth is highlighted.

Adelomou et al. \cite{adelomou2020using} address personnel scheduling. They look at the challenges faced by social workers in generating optimal visiting schedules for patients. The problem involves a combination of routing and planning issues. The article suggest the use of the Parameterised Quantum Circuits (PQC) and the VQE algorithm in solving the problem. It explores the potential of PQC as a basis for Machine Learning (ML) and delves into the formulation of the social worker problem for quantum computing. The proposed methodology aims to offer an efficient and intelligent system, capable of learning by analogy and avoiding the need for recalculating from scratch. Experimental results suggest the effectiveness of the Quantum Cases-Based Reasoning (QCBR) approach, demonstrating potential benefits for optimising worker schedules. The article concludes by emphasising the broader applicability of the proposed formulation to solve various planning, scheduling, and routing problems. It envisions the creation of hybrid and intelligent systems that leverage quantum resources efficiently, reducing costs and laying the foundation for generalised problem-solving in the field. 

Last, Krol et al. \cite{krol2024qiss} show the design and implementation of a quantum algorithm for industrial shift scheduling (QISS), which uses Grover's adaptive search to tackle a common and important class of valuable, real-world combinatorial optimization problems. They give an explicit circuit construction of the Grover's oracle, incorporating the multiple constraints present in the problem, and detail the corresponding logical-level resource requirements. Further, they simulate the application of QISS to specific small-scale problem instances to corroborate the performance of the algorithm, and provide an open-source repository.

\section{Conclusions and Recommendations}
The total overview of all the literature discussed in the previous section can be found in Table \ref{tab:my_label1} and Table \ref{tab:my_label2}. Here the paradigm, the main topic and the discussed problem are listed. Also, in these tables is indicated whether the solution is a pure quantum solution or a hybrid quantum solution. With hybrid we mean `Horizontal Hybrid Quantum Computing', following the classification of \cite{phillipson2023classification}. This means that the algorithm requires both a quantum computer and a classical computer to perform an algorithm, in sequential or parallel order. As we can see in the tables, most of the current solutions use hybrid approaches. This is as expected, from two thoughts. First, the current generation of quantum computers are still small, meaning that solving real world use cases, which is what most articles in the literature aim for, is limited and classical computers are needed to perform at least part of the algorithm. Next, for many optimisation problems a hybrid approach, using quantum computers as part of a (meta-) heuristic approach is expected to last, where also classical algorithms break down the calculation pipeline and the QC is not expected to solve NP-hard problems \cite{aaronson2008limits}. These meta-heuristic approaches, where a QC is combined with a classical counterpart report the best performances.
An other aspect that is clearly visible in this overview is that the majority of publications use QA as main quantum paradigm. This follows by the relative high number of qubits and larger size of problems this technology allows for. Also the hybrid approaches, as offered by D-Wave in their full stack approach, help with the uptake here. The QAOA and other VQE approaches on the gate-based device are the second most used approach, however they are still very much limited in performance. More problem specific and machine learning approaches are still limited in number. There is quite a bit of room for experimentation here.
When we look at the topics and problems covered, we see most attention to the classical areas of routing and scheduling. Especially in predictions there is again room for further research, potentially for QML approaches.

Generally, we can say that we endorse the conclusions reached in the work of Osaba et al. \cite{osaba2022systematic}. Where we looked more general than only VRP and TSP work, also we can conclude that the majority of the papers conclude that current hardware is in the early stage: researchers face small number of qubits and inherent limitations such as noise and decoherence, and challenges in tailoring problems to hardware capacities. This leads to limitations in addressing complex formulations. Only a small number of papers indicate that their hybrid approach beats current classical approaches. Most of the work indicate the \textbf{\textit{potential}} advantages and benefits of their approach.

\section*{Acknowledgements}
This publication is part of the project Divide and Quantum `D\&Q' NWA.1389.20.241 of the program `NWA-ORC', which is partly funded by the Dutch Research Council (NWO).
The author likes to thank Juan Boschero and Stan van der Linde for their valuable input.

\begin{table}[h!]
    \centering
    \begin{tabular}{l|l|l|l|l}
Work	& Paradigm & Topic & Problem & Nature \\ \hline
Ajagekar \cite{ajagekar2020quantum2} & QA - QUBO & Routing & VRP & hybrid\\
Alsaiyari \cite{alsaiyari2023variational}&GBC - QAOA/VQE & Routing & VRP & hybrid\\
Atchade-Adelomou \cite{atchade2021qrobot} & QA and GBC - QUBO & Routing & picking & hybrid\\
Azad \cite{azad2022solving} & GBC - QAOA & Routing & VRP & hybrid\\
Azzaoui \cite{azzaoui2021quantum} & GBC - QAOA & Routing & TSP & hybrid\\
Bentley\cite{bentley2022quantum} & GBC - QAOA & Routing & CVRP & hybrid\\
Borowski \cite{borowski2020new} & QA - QUBO & Routing & VRP/CVRP & full and hybrid\\
Bourreau \cite{bourreau2023indirect} & GBC - QAOA & Routing & TSP & hybrid\\
Cattelan \cite{cattelan2024modeling} & QA - QUBO & Routing & RPP & hybrid\\
Correll \cite{correll2023quantum} & GBC - QML & Routing & VRP & hybrid\\ 
Dixit \cite{dixit2023quantum2} & QA - QUBO & Routing & STDSP & full\\
Feld \cite{feld2019hybrid} & QA - QUBO & Routing & CVRP & hybrid\\
Fitzek \cite{fitzek2021applying}& GBC - QAOA & Routing & HVRP & hybrid\\
Fitzek \cite{fitzek2024applying}& GBC - QAOA & Routing & HVRP & hybrid \\
Harikrishnakumar \cite{harikrishnakumar2020quantum}& QA - QUBO & Routing & (D-)MDCVRP & full\\ 
Herzog \cite{herzog2024improving} & GBC - QAOA & Routing & CVRP & hybrid\\
Harwood \cite{harwood2021formulating} & GBC - QAOA & Routing & VRPTW & hybrid\\
Irie \cite{irie2019quantum} & QA - QUBO & Routing & CVRPTW & hybrid\\
Kanai \cite{kanai2024annealing} & QA - QUBO & Routing & CVRP & hybrid\\
Khan \cite{khan2024optimizing} &  GBC -Grover's/QAOA & Routing & FRO & hybrid \\
Khumalo \cite{khumalo2022investigation} &GBC - QAOA/VQE & Routing & TSP & hybrid\\
Le \cite{le2023quantum}& QA - QUBO & Routing & sTSP & hybrid\\
Leonidas \cite{leonidasqubit}& GBC - QAOA & Routing & VRPTW & hybrid\\
Li \cite{li2024quantum} &  GBC - VQA & Routing & VRP & hybrid \\
Liu \cite{liu2024quantum} & QA - QUBO & Routing & TSP & hybrid \\
Lo \cite{lo2021genetic} & QRNG & Routing & VRP & hybrid\\
Makhanov \cite{makhanov2023quantum}& GBC - Grover's & Routing & SPP & hybrid\\ 
Makhanov \cite{makhanov2024quantum} & GBC - Grover's & Routing & FRO & hybrid\\
Mario \cite{mario2024quantum} & QA - QUBO & Routing & CVRP & hybrid\\
Marsoit \cite{marsoit2021quantum} & QA - QUBO & Routing & CVRP & hybrid\\
Masuda \cite{masuda85optimisation}& CA - QUBO & Routing & TDVRPTW & full\\
Mohanty \cite{mohanty2023analysis} & GBC - VQE & Routing & VRP & hybrid\\
Mohanty \cite{mohanty2024solving} & GBC - QSVM & Routing & VRP & hybrid\\
Neukart \cite{neukart2017traffic} & QA - QUBO & Routing & Flow opt. & hybrid\\
Osaba \cite{osaba2024solving} &	QA - QUBO & Routing & CVRP & hybrid\\
Palackal \cite{palackal2023quantum}& GBC - QAOA/VQE & Routing & TSP & hybrid\\
Palmieri \cite{palmieriquantum} & QA and GBC - QUBO & Routing & CVRP & hybrid\\
Papalitsas \cite{papalitsas2019qubo} & QA - QUBO & Routing & TSPTW & full\\ 
Phillipson \cite{phillipson2021multimodal} & QA - QUBO & Routing & Flow opt. & hybrid\\
Poggel \cite{poggel2023recommending} & Generic & Routing & CVRP & hybrid\\
Qiu \cite{qiu2024novel} & GBC - QACO & Routing & TSP & hybrid\\
Ramezani \cite{ramezani2024reducing} & GBC - QAOA & Routing & TSP & hybrid\\
Rana \cite{rana2022quantum} & QA and GBC - QUBO & Routing & CVRP & hybrid\\
Rosendo \cite{rosendo2024quantum} & QA and GBC - QUBO & Routing & CVRP & hybrid\\
Sadashivan \cite{sadashivan2024flexible} & QA - QUBO & Routing & CVRP & hybrid \\
Salehi \cite{salehi2022unconstrained} & QA - QUBO & Routing & TSPTW & full\\ 
Sales \cite{sales2023adiabatic} & QA - QUBO & Routing & CVRP & full and hybrid\\ 

    \end{tabular}
    \caption{Overview - Part I}
    \label{tab:my_label1}
\end{table}
\newpage
\begin{table}[h!]
    \centering
    \begin{tabular}{l|l|l|l|l}
Work	& Paradigm & Topic & Problem & Nature \\ \hline
Salloum \cite{salloum2024mini} & QA - QUBO & Routing & TFO & hybrid\\
Salloum \cite{salloum2024quantum} & QA - QUBO & Routing & TFO & hybrid\\
Sanches \cite{sanches2022short} & GBC - RL & Routing & CVRP & hybrid\\
Sato \cite{sato2024circuit} & GBC - Quantum Search & Routing & TSP & full\\
Sinno \cite{sinno2023performance} & QA - QUBO & Routing & CVRP & hybrid\\ 
Spyridis \cite{spyridis2023variational}& GBC - QAOA & Routing & m-TSP& hybrid\\
Srinivasan \cite{srinivasan2018efficient} & GBC - QPE & Routing & TSP& hybrid\\
Tejani \cite{tejani2024quantum} & GBC - QPE & Routing & BTSP & hybrid\\
Warren \cite{warren2020solving} & QA - QUBO & Routing & TSP & full and hybrid\\ 
Weinberg \cite{weinberg2023supply} & QA - QUBO & Routing & CVRP & hybrid\\ 
Wesolowski \cite{wesolowski2024advances} & GBC & Routing & SPSP & full\\
Xie \cite{xie2023feasibility}  & GBC - QAOA & Routing & CVRP & hybrid\\
Xu \cite{xu2024quantum} & GBC - QML & Routing & CVRP & hybrid\\
Yarkoni \cite{yarkoni2021solving} &QA - QUBO & Routing & SRP & hybrid\\ \hline
Bu{\ss} \cite{bussnear} & GBC - QAOA & Network Design & FLP & hybrid\\
Chiscop \cite{chiscop2020hybrid} & QA - QUBO & Network Design & MSLSCP & hybrid\\
Choo \cite{chooinvestigating} & QA - QUBO & Network Design & FLP & hybrid\\
Ding \cite{ding2021implementation} & QA -QUBO & Network Design & NDP & hybrid\\
Ding \cite{ding2019logistic}& QA -QUBO & Network Design & NDP & full\\
Dixit \cite{dixit2023quantum}& QA -QUBO & Network Design & NDP & hybrid\\
Gabbassov \cite{gabbassov2022transit} & QA - QUBO & Network Design & UTP & hybrid\\
Giraldo \cite{giraldo2022using}& QA and GBC - QUBO & Network Design & MCLP & hybrid\\
Guo \cite{guo2020quantum} & QA - QUBO & Network Design & FLP & hybrid\\
Khumalo \cite{khumalo2022investigation} &GBC - QAOA/VQE & Network Design & FLP & hybrid\\
Klar \cite{klar2022quantum} & QA - QUBO & Network Design & FLP & hybrid\\
Mahasinghe \cite{mahasinghe2021qubo} & QA - QUBO & Network Design & FLP & -\\
Malviya \cite{malviya2023logistics}& QA - QUBO & Network Design & NDP & hybrid\\
Malviya \cite{malviya2024redesign} & GBC - QAOA & Network Design & NDP & full\\
Radvand \cite{radvand2024quantum} & GBC -Grover's & Network Design & CSLP & full\\
Sakib \cite{sakib2024quantum} & QA - QUBO & Network Design & CSLP & hybrid\\
Satori \cite{satori2023quantum} & QA - QUBO & Network Design & LAP & hybrid\\
Tosum \cite{tosun2022new} & QA - QUBO & Network Design & QAP & -\\
Wang \cite{wang2024variational} & GBC - QAOA & Network Design & FLP &hybrid\\ \hline
Martins \cite{martins2020applying} & QA - QUBO & Fleet opt. & TAP & hybrid\\
Vikstaal \cite{vikstaal2020applying} & GBC - QAOA & Fleet opt. & TAP & hybrid\\
Willsch \cite{willsch2022benchmarking} & QA - QUBO & Fleet opt. & TAP & hybrid\\ \hline
De Andoin \cite{de2022comparative}& QA - QUBO & Cargo &  BPP & hybrid\\
De Andoin \cite{de2022hybrid} & QA - QUBO & Cargo &  BPP & hybrid\\
Ardelian \cite{ardelean2024hybrid} & GBC - QML & Cargo & QKP & hybrid \\
Benson \cite{benson2023cqm}& QA - CQM & Cargo & KP & hybrid\\
Bontekoe \cite{bontekoe2023translating}& QA - QUBO & Cargo & QKP & full\\
Bozejko \cite{bozejko2024optimal} &	QA - QUBO & Cargo & QKP & hybrid\\
Cellini \cite{cellini2023qal}& QA - QUBO & Cargo &  BPP & full\\
Cellini \cite{cellini2024qal} & QA - QUBO & Cargo & BPP & hybrid\\
Christiansen \cite{christiansen2024quantum} & GBC - QAOA & Cargo & QKP & hybrid\\
Cui \cite{cui2024hybrid} & GBC - Grover's & Cargo & QKP & hybrid\\
Gatti \cite{gatti2024qubo} & QA - QUBO & Cargo & BPP & hybrid\\
    \end{tabular}
    \caption{Overview - Part II}
    \label{tab:my_label2}
\end{table}
\newpage
\begin{table}[h!]
    \centering
    \begin{tabular}{l|l|l|l|l}
Work	& Paradigm & Topic & Problem & Nature \\ \hline
Matt \cite{matt2024heuristic} & QAOA & Cargo & BPP & hybrid \\
Nayak \cite{nayak2022quantum} & QA - QUBO & Cargo & ALO & hybrid\\
Pilon \cite{pilon2021aircraft} & QA - QUBO & Cargo &  ALO & full\\
Romero  \cite{romero2023solving,v2023hybrid}  & QA - CQM & Cargo &  BPP & hybrid\\
Shirai \cite{shirai2024post} & GBC - VQA & Cargo & QKP & hybrid\\
Sotelo \cite{hernandez2020application,sotelo2021determination} & GBC - VQE & Cargo & ALO & hybrid\\
Van Dam \cite{van2021quantum} & GBC - QAOA & Cargo & KP & hybrid\\ \hline

Gutta \cite{gutta2024ai} &GBC - QML & Prediction & Forecasting & hybrid \\
Jahin \cite{jahin2023qamplifynet}&GBC - QML & Prediction & Prediction & hybrid\\
Jiang \cite{jiang2022quantum} & GBC - HHL/VQE & Prediction & Inventory & hybrid\\
Koushik \cite{koushik2024enhancing}	&GBC - QML & Prediction & Maintenance & hybrid\\
Sehrawat \cite{sehrawat2024predicting} &GBC - QML & Prediction & Forecasting & hybrid\\ \hline

Adelomou \cite{adelomou2020using} &GBC - VQE& Scheduling & Personnel & hybrid\\
Ajagekar \cite{ajagekar2020quantum2} & QA - QUBO & Scheduling & JSP & hybrid\\
Ajagekar \cite{ajagekar2020quantum2} & QA - QUBO & Scheduling & MCFP & hybrid\\
Amaro \cite{amaro2022case} & GBC - VQE& Scheduling & JSP & hybrid\\
Bernreuther \cite{bernreuther2024integrating} & GBC - QAOA & Scheduling & Scheduling & hybrid \\
Carugno \cite{carugno2022evaluating} & QA - QUBO & Scheduling & JSP & hybrid\\ 
Denkena \cite{denkena2021quantum} & QA - QUBO & Scheduling & FJSP & hybrid\\ 
Grange \cite{grange2024design} & GBC - QAOA & Scheduling & Scheduling & hybrid \\
Huang \cite{huang2022paint} & QA - QUBO & Scheduling & JSP & hybrid\\
Krol \cite{krol2024qiss} & GBC - Grover's & Scheduling & Scheduling & full\\
Kurowski \cite{kurowski2020hybrid} & QA - QUBO & Scheduling & JSP & hybrid\\ 
Mohammadbagherpoor \cite{mohammadbagherpoor2021exploring}& QA - VQE & Scheduling & QAP & hybrid\\
Mori \cite{mori2022replanning} & GBC and QA - QAOA & Scheduling & ARP & hybrid\\ 
Permin \cite{permin2022simple} & GBC - Grover & Scheduling & JSP & hybrid\\
Riandari \cite{riandari2021quantum}&GBC - QAOA & Scheduling & Production opt. & hybrid\\ 
Scherer \cite{scherer2021oncall} &GBC - Grover & Scheduling & JSP & full\\
Streif \cite{streif2021beating}& GBC - QAOA & Scheduling & JSP & hybrid\\
Schworm \cite{schworm2023solving} & QA - QUBO & Scheduling & JSP & hybrid\\
Schworm \cite{schworm2024evaluation}&QA - QUBO&Scheduling&JSP&hybrid\\
Schworm \cite{schworm2024multi}&QA - QUBO&Scheduling&JSP&hybrid\\
Venturelli \cite{venturelli2015quantum}& QA - QUBO & Scheduling & JSP & full\\
Windmann \cite{windmann2024quantum}&GBC - QAOA&Scheduling&JSP&hybrid\\
Yarkoni \cite{yarkoni2021multi} & QA - QUBO & Scheduling & JSP & hybrid\\
    \end{tabular}
    \caption{Overview - Part III}
    \label{tab:my_label3}
\end{table}

\begin{tabular}{@{}ll}
ALO & Aircraft Loading Optimisation\\
AP & Assignment Problem\\
ARP & Aircraft Recovery Problem\\
AS/RS & Automated Storage and Retrieval System\\
BPP & Bin Packing Problems\\
CA & Classical annealing \\
CSLP & Charging Station Location Problem\\
CQM & Constrained Quadratic Model\\
CVRP&Capacitated Vehicle Routing Problem\\
FJSP & Flexible Job Shop scheduling Problem\\
FLP & Factory Layout Planning\\
GBC & Gate Based Computer/Computing\\
HPC & High Performance Computing \\
HVRP & Heterogeneous Vehicle Routing Problem\\
JSP & Job Shop Problem \\
KP & Knapsack Problem \\
LAP & Location Assignment Problem \\
LND & Logistic Network Design\\
MCFP & Manufacturing Cell Formation Problem\\
(D-)MDCVRP & (Dynamic) Multi-Depot Capacitated Vehicle Routing Problem\\
MIP & Mixed-Integer Programming\\
MCLP & Maximal Covering Location Problem\\
MDKP & Multi-Dimensional Knapsack Problem\\
MSLSCP & Multi-Service Location Set Covering Problem\\
NDP & Network Design Problem\\
NISQ & Noisy Intermediate-Scale Quantum (devices)\\
QA & Quantum Annealing \\
QAOA &  Quantum Approximate Optimisation Algorithm \\
QAP & Quadratic Assignment Problem \\
QKP & Quadratic Knapsack Problem \\
QML & Quantum Machine Learning \\
QPE & Quantum Phase Estimation \\
QPU & Quantum Processing Unit \\
QRNG & Quantum Random Number Generator\\
QUBO & Quadratic Unconstrained Binary Optimisation problem\\
RL & Reinforcement Learning \\
SIC & Spatial Interaction Coverage\\
SPP & Shortest Path Problem\\
SRP & shipment rerouting problem \\
STDSP & Stochastic Time Dependent Shortest Path problem\\
TAP & Tail Assignment Problem \\
TDVRPTW & Time-Dependent Vehicle Routing Problem with Time Windows\\
TSP & Travelling Salesman Problem\\
TSPTW & Travelling Salesman Problem with Time Windows\\ 
m-TSP & multiple Travelling Salesman Problem\\
sTSP & selective travelling Salesman Problem\\
UTP & urban transit planning\\
VQE & Variational Quantum Eigensolver\\
VRP &Vehicle Routing Problem\\
VRPTW & Vehicle Routing Problem with Time Windows\\
\end{tabular}

\bibliographystyle{splncs04}
\bibliography{literature.bib}
\end{document}